\documentclass[useAMS,usenatbib]{mn2e}

\usepackage{graphicx,amssymb,hyperref}
\usepackage[fleqn]{amsmath}  
\usepackage{multirow}
\usepackage{graphicx} 
\usepackage{color}

\usepackage{graphicx,amssymb,multirow}
\usepackage[fleqn]{amsmath}

\def\simlt{\lower.5ex\hbox{$\; \buildrel < \over \sim \;$}}
\def\simgt{\lower.5ex\hbox{$\; \buildrel > \over \sim \;$}}

\def\ie{{\it i.e.}}
\def\eg{{\it e.g.}}

\newcommand{\be}{\begin{equation}}
\newcommand{\ee}{\end{equation}}
\newcommand{\ba}{\begin{eqnarray}}
\newcommand{\ea}{\end{eqnarray}}

\usepackage{graphicx} 

\title[Dark matter in galaxy cluster Abell 3827]{The behaviour of dark matter associated with 4 bright cluster galaxies in the 10\,kpc core of Abell 3827}
\author[R.\ Massey et al.]{Richard Massey$^{1,2}$\thanks{e-mail: {\tt r.j.massey@durham.ac.uk}},
Liliya Williams$^{3}$,
Renske Smit$^{2}$,
Mark Swinbank$^{2}$, 
Thomas D.\ \newauthor Kitching$^{4}$,
David Harvey$^{5}$,
Mathilde Jauzac$^{1,6}$, 
Holger Israel$^{1}$,
Douglas Clowe$^{7}$, \newauthor
Alastair Edge$^{2}$,
Matt Hilton$^{6}$,
Eric Jullo$^{8}$,
Adrienne Leonard$^{9}$, 
Jori Liesenborgs$^{10}$, \newauthor
Julian Merten$^{11,12}$, 
Irshad Mohammed$^{13}$,
Daisuke Nagai$^{14}$, 
Johan Richard$^{15}$, \newauthor
Andrew Robertson,$^{\!\!2}$ 
Prasenjit Saha,$^{\!\!13}$
Rebecca Santana,$^{\!\!7}$
John Stott$^{2}$ \& 
Eric Tittley$^{16}$\\
$^{1}$ Institute for Computational Cosmology, Durham University, South Road, Durham DH1 3LE, UK\\ 
$^{2}$ Centre for Extragalactic Astronomy, Durham University, South Road, Durham DH1 3LE, UK\\
$^{3}$ School of Physics \& Astronomy, University of Minnesota, 116 Church Street SE, Minneapolis, MN 55455, USA\\
$^{4}$ Mullard Space Science Laboratory, University College London, Holmbury St Mary, Dorking, Surrey RH5 6NT, UK\\
$^{5}$ \'Ecole Polytechnique F\'ed\'erale de Lausanne, 51 Chemin des Maillettes, Observatoire de Sauverny, Versoix, CH-1290 Switzerland\\
$^{6}$ Astrophysics and Cosmology Research Unit, School of Mathematical Sciences, University of KwaZulu-Natal, Durban 4041, South Africa\\
$^{7}$ Department of Physics and Astronomy, Ohio University, 251B Clippinger Labs, Athens, OH 45701, USA\\
$^{8}$ Aix Marseille Universit\'e, CNRS, LAM (Laboratoire d'Astrophysique de Marseille), UMR 7326, 13388, Marseille, France\\
$^{9}$ University College London, Gower Street, London WC1E 6BT, UK\\
$^{10}$ Expertisecentrum voor Digitale Media, Universiteit Hasselt, Wetenschapspark 2, B-3590, Diepenbeek, Belgium\\
$^{11}$ Jet Propulsion Laboratory, California Institute of Technology, 4800 Oak Grove Drive, Pasadena, CA 91109, USA\\
$^{12}$ California Institute of Technology, MC 249-17, Pasadena, CA 91125, USA\\
$^{13}$ Physik-Institut, University of Z\"urich, Winterthurerstrasse 190, 8057 Z\"urich, Switzerland\\
$^{14}$ Department of Physics, Yale University, New Haven, CT 06520, USA\\
$^{15}$ Observatoire de Lyon, Universit\'e Lyon 1, 9 Avenue Charles Andr\'e, 69561 Saint Genis Laval Cedex, France\\
$^{16}$ Royal Observatory, Blackford Hill, Edinburgh EH9 3HJ, UK
}
\begin{document}
\date{Accepted 2015 March 02. Received 2015 March 02; in original form 2014 December 04}

\pagerange{\pageref{firstpage}--\pageref{lastpage}} \pubyear{2014}

\maketitle

\label{firstpage}

\begin{abstract}

Galaxy cluster Abell 3827 hosts the stellar remnants of four almost equally bright elliptical galaxies within a core of radius 10\,kpc.
Such corrugation of the stellar distribution is very rare, and suggests recent formation by several simultaneous mergers.  
We map the distribution of associated dark matter, using new {\sl Hubble Space Telescope} imaging and\,{\sl VLT}/{\sl MUSE} integral field spectroscopy of a gravitationally lensed system threaded through the cluster core.
We find that each of the central galaxies retains a dark matter halo, but that (at least) one of these is spatially offset from its stars.
The best-constrained offset is $1.62^{+0.47}_{-0.49}$\,kpc, where the 68\% confidence limit includes both statistical error and systematic biases in mass modelling.  
Such offsets are not seen in field galaxies, but are predicted during the long infall to a cluster, if dark matter self-interactions generate an extra drag force.
With such a small physical separation, it is difficult to definitively rule out astrophysical effects operating exclusively in dense cluster core environments -- but if interpreted solely as evidence for self-interacting dark matter, this offset implies a cross-section $\sigma_\mathrm{DM}/m\sim\!(1.7\pm0.7)\!\times\!10^{-4}$\,cm$^2$/g $\times(t_\mathrm{infall}/10^{9}\,{\rm yrs})^{-2}$, where $t_\mathrm{infall}$ is the infall duration.
\end{abstract}


~\newpage ~\newpage 
\section{Introduction}

Many lines of evidence now agree that most mass in the Universe is in the form of dark matter, which interacts mainly via the force of gravity.
The identity and detailed phenomenology of dark matter remain poorly understood.
However, its gravitational attraction pulls low-mass systems into a series of hierarchical mergers 
through which dark and ordinary matter are gradually assembled into giant clusters of galaxies
\citep{dav85}.

The typically smooth distribution of light in galaxy clusters visible today shows that merging systems have their gas content efficiently removed into the intra-cluster medium by ram pressure stripping, even while they pass the virial radius \citep{smi10,wu12}.
The longevity of accompanying dark matter is less well understood, but the timescale for its dissipation is a key ingredient in semi-analytic models of structure formation \citep{dar10}.
Full numerical simulations predict that the dark matter is eventually smoothed \citep{gao04,nag05,bah12}, but disagree about the timescale and the radius/orbits on which stripping occurs \citep{die07,pen08,wet09}.
Observations have shown that, as $L$* galaxies enter a galaxy cluster from the field, tidal gravitational stripping of their dark matter \citep{man06,par07,gil13} reduces their masses by $\sim\!10^{13}M_\odot$ to $\sim\!10^{12}M_\odot$ from a radius of 5\,Mpc to 1\,Mpc \citep{lim07,lim12,nat09}.
This stripping occurs at a rate consistent with simulations, but has not been followed to the central tens of kiloparsecs, which is where the predictions of simulations disagree. 

Mergers of dark matter substructures into a massive galaxy cluster also reveal the fundamental properties of dark matter particles. 
The different non-gravitational forces acting on dark matter and standard model particles have been highlighted most visibly in collisions like the `Bullet Cluster' 1E0657-56 \citep{clo04,clo06,bra06}, Abell~520 \citep{mah07,clo12,jee14}, MACSJ0025-12 \citep{bra08}, Abell~2744 \citep{mer11} and DLSCL J0916.2+2951 \citep{daw12}. 
Infalling gas (of standard model particles) is subject to ram pressure and tends to lag behind non-interacting dark matter \citep{you11}.
Measurements of this lag yielded an upper limit on dark matter's self-interaction cross section $\sigma/m\!<\!1.2$\, cm$^2\!/$g if the particle momentum exchange is isotropic, or $\sigma/m\!<\!0.7$\, cm$^2\!/$g if it is directional \citep{ran08,kah14}. 
More interestingly still, if dark matter has (even a small) non-zero self-interaction cross-section, infalling dark matter will eventually lag behind old stars \citep{mkn11,ws11,kah14,har13,har14}.
Self-interactions confined within the dark sector can potentially have much larger cross-sections than those between dark matter and standard model particles, which are constrained by collider and direct detection experiments \citep[\eg][]{peterreview}.

The galaxy cluster Abell~3827 (RA=$22$h\,$01\arcmin$\,$49\farcs1$, Dec=$-59^\circ$\,$57\arcmin$\,$15\arcsec$, $z$=0.099, X-ray luminosity $L_X=8\times10^{44}$\,erg/s in the $0.1$--$2.4$\,keV band, \citealt{depla07}) is particularly interesting for substructure studies because it hosts the remnant stellar nuclei of {\it four} bright elliptical galaxies within the central 10\,kpc.
Such {\it corrugation of the stellar distribution} is very rare: only Abell~2261 \citep{coe12,pos12} and MACSJ0717 \citep{lim12,jau12} are even comparably corrugated.
All three clusters are still forming, through several simultaneous mergers -- and can be used to investigate the late-stage dissipation of dark matter infalling {\it through the same environment}.
Moreover, Abell~3827 has a unique strong gravitational lens system threaded between its multiple central galaxies \citep{car10}.
This enables the distribution of its otherwise invisible dark matter to be mapped \citep[for reviews of gravitational lensing, see][]{bs01,refrev,hoerev,mrev,barrev,knerev}.
The cluster even lies within the optimum redshift range $0.05\!<\!z\!<\!0.1$ to measure small physical separations between dark and ordinary matter \citep{mkn11}.

Indeed, ground-based imaging \citep{ws11,moh14} suggests that dark matter associated with one of the central galaxies in Abell~3827 (the one where its position is best constrained, `nucleus' N.1) is offset by $\sim\!3\arcsec$ (6\,kpc) from the stars.
Such offsets are not seen in isolated field galaxies \citep[\eg][]{slacs3,slacs4}.
Interpreting the offset via a model in which $t_\mathrm{infall}$ is the time since infall, implies a lower limit of $\sigma/m\!>\!4.5\times10^{-6}(t_\mathrm{infall}/10^{10}{\mathrm yr})^{-2}$\,cm$^2\!/$g.
This is potentially the first detection of non-gravitational forces acting on dark matter.

In this paper, we present new {\sl Hubble Space Telescope} ({\sl HST}) imaging and {\sl Very Large Telescope} ({\sl VLT}) integral-field spectroscopy to hone measurements of the dark matter distribution.
We describe the new data in section~\ref{sec:data}, and our mapping of visible light/dark matter in section~\ref{sec:analysis}.
We describe our results in section~\ref{sec:results}, and discuss their implications in section~\ref{sec:discussion}.
We conclude in section~\ref{sec:conc}.
Throughout this paper, we quote magnitudes in the AB system and adopt a cosmological model with $\Omega_\mathrm{M}=0.3$, $\Omega_\Lambda=0.7$ and $H_0=70$\,km/s/Mpc, in which $1\arcsec$ corresponds to $1.828$\,kpc at the redshift of the cluster.

\section{Data} \label{sec:data}

\begin{figure*}
\begin{center}
\includegraphics[width=0.95\textwidth]{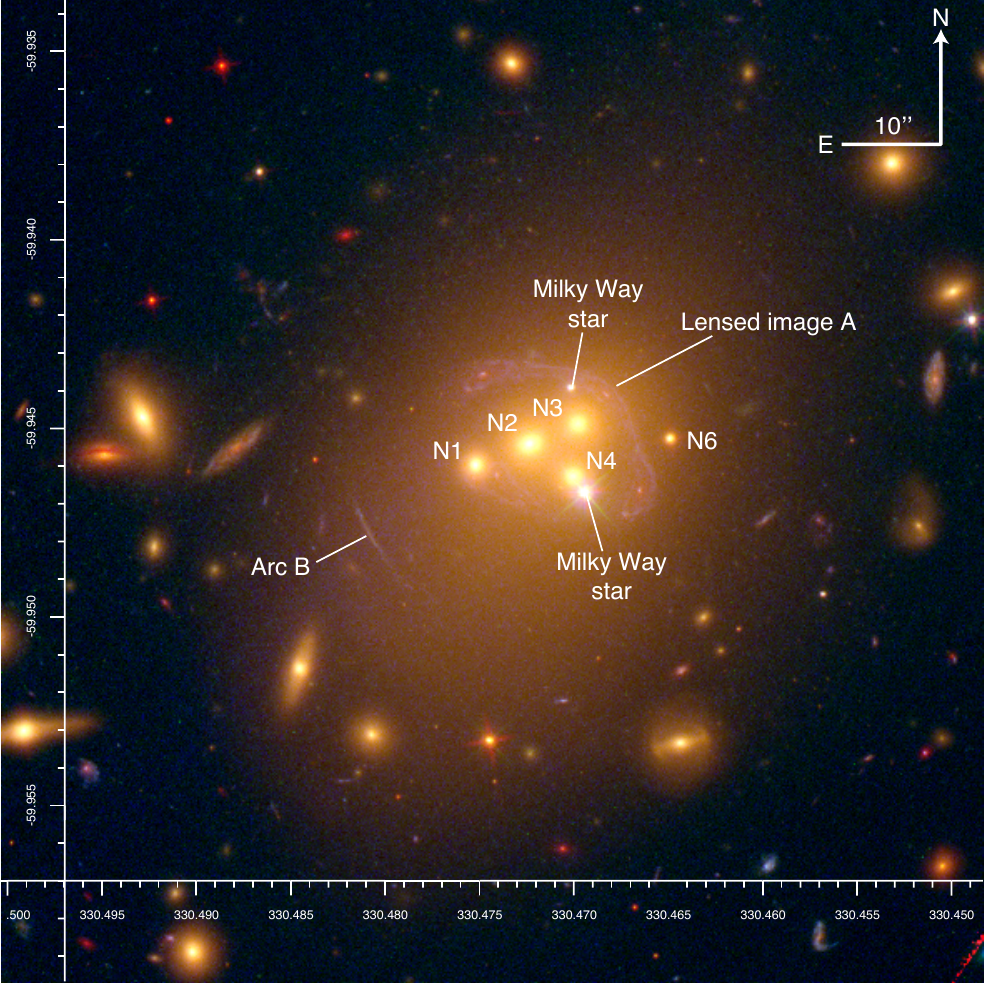}\\
\caption{{\sl Hubble Space Telescope} image of galaxy cluster Abell~3827, showing the F160W (red), F606W (green) and F336W (blue) bands. 
The colour scale is logarithmic.
Labels show the four bright (plus one faint) central galaxies, foreground stars and background lensed galaxies. 
An object previously referred to as N.5 is actually a star.
All these identifications are confirmed spectroscopically.} \label{fig:image}
\end{center}
\end{figure*}

\subsection{HST imaging}

We imaged the galaxy cluster Abell 3827 using the {\sl Hubble Space Telescope (HST)}, programme GO-12817. 
Observations with the {\sl Advanced Camera for Surveys (ACS)}/{\sl Wide Field Channel (WFC)} during October 2013 consisted of 5244\,s in optical band F814W (in the core, with half that depth across a wider area) and 2452\,s in optical band F606W.
Observations with the {\sl Wide Field Camera 3 (WFC3)} in August 2013 consisted of 5871\,s in UV band F336W and 2212\,s in near-IR band F160W.

The raw data exhibited spurious trailing due to Charge Transfer Inefficiency in imaging detectors damaged by radiation. 
We corrected this trailing using the software of \citet{m10,m14} for the {\sl ACS}/{\sl WFC} and \citet{and10,and14} for {\sl WFC3}/{\sl UVIS}.
Subsequent data reduction then followed the standard procedures of {\sc calacs} v2012.2 \citep{calacs} and {\sc calwf3} v2.7 \citep{calwf3}.
We stacked individual exposures using {\sc drizzle} \citep{drizzle} with a Gaussian convolution kernel and parameter {\sc pixfrac}=0.8, then aligned the different observations into the common coordinate system of the F814W data using {\sc tweakback}.
Figure~\ref{fig:image} shows a multicolour image of the cluster core. 

\subsection{VLT spectroscopy} \label{sec:IFUreduction}

We first obtained spectroscopy across the cluster core using the 
{\sl VLT}/{\sl VIMOS} integrated field unit \citep[IFU;][]{vimos,lef13},
programme 093.A-0237.  Total exposure times were 6\,hours in the
HR-blue filter (spanning a wavelength range $370$--$535$\,nm with
spectral resolution $\lambda/\Delta\lambda\!=\!200$) during July 2014
and 5\,hours in the MR-orange filter ($490$--$1015$\,nm, with
$\lambda/\Delta\lambda\!=\!1100$) during August 2014.  All
observations were obtained in photometric conditions and
$<\!0\farcs6$ seeing, using the $27\arcsec\!\times27\arcsec$ field
of view; in this configuration, each pixel is $0\farcs66$ on a side.

Since the cluster core is high surface brightness across the entire {\sl VIMOS}
field of view, we interspersed every three exposures on target with
one offset by $\sim\!2\arcmin$ to record (and subtract) the sky
background.  The three on-source exposures were dithered by
$1\farcs3$ (2\,pixels) to account for bad fibres and cosmetics.

To reduce the raw data we used the {\sl VIMOS} {\sc esorex} pipeline, which
extracts the fibres, wavelength calibrates and flatfields the data,
and forms the data cube.  We used the temporally adjacent sky exposure
to perform sky subtraction (on a quadrant-by-quadrant basis), then
mosaiced all of the exposures using a clipped average (using the
bright stars to measure the relative offsets between cubes).  We
constructed (wavelength collapsed) continuum images from the cubes and
aligned the cube to the {\sl HST} imaging, then extracted spectra for
each of the continuum sources.  To search for emission from the strong
lensing features, we applied a mask to the cube and extracted both
one- and two- dimensional spectra.  Due to the different resolutions
of the HR-blue and MR-orange observations, we analysed the final two
data cubes separately, but overall they provide a continuous wavelength
coverage from $370$ to $1015$\,nm.
This is perfectly sufficient for our analysis of the cluster light.

The lensed galaxy threaded through the cluster core (labelled A in figure~\ref{fig:image}) was originally
identified in {\sl Gemini} imaging by \citet{car10}, who also used
long-slit {\sl Gemini} spectroscopy to obtain a redshift
$z\!=\!0.204$.  However, our IFU spectroscopy did not confirm
this. We instead found only one bright emission line at 835.5\,nm, which 
is not associated with lines at the foreground cluster redshift and whose 2D
morphology traces the lensed image.  The emission line could have been H$\alpha$ at
$z\approx0.27$ or [O{\sc ii}] at $z\approx1.24$ --- but the low
resolution and signal-to-noise of the {\sl VIMOS} data precluded robust
identification (the lack of other features, such as [N{\sc
    ii}]\,$\lambda$6853 for $z$\,=\,0.27 or [O{\sc
    iii}]\,$\lambda$5007 for $z$\,=\,1.24 may have been due to low
metallicity and\,/\,or low signal-to-noise in the lines).  Moreover,
the flat-field (in)stability in {\sl VIMOS} data left 
strong residuals after subtracting foreground emission from the cluster galaxies.
In particular, this made it difficult to robustly determine the morphology of the arc 
near N.1, and --- as we shall see in section~\ref{sec:resultsdm} --- this provides the most 
diagnostic power in the lens modelling.

To confirm the redshift of lensed system A, and to measure its
morphology around N.1, we observed the cluster core
with the {\sl VLT} {\sl Multi-Unit Spectroscopic Explorer (MUSE)}
IFU spectrograph \citep{muse} during Director's Discretionary Time in 
December 2014, program 294.A-5014.  
The awarded exposure time of 1\,hour 
was split in to $3\times1200$\,s exposures, which
were dithered by $\sim$10$''$ to account for cosmic rays and defects.
These observations were taken in dark time, $<$0.7$''$ $V$-band seeing
and good atmospheric transparency.  
We stacked these with an extra $1200$\,s exposure that was taken in good seeing during twilight, which marginally improves the signal to noise.
{\sl MUSE} has a larger $1\arcmin\!\times1\arcmin$  field of
view and excellent flatfield stability, so no extra sky exposures
were required.  The data were reduced using v1.0 of the {\sc esorex}
pipeline, which extracts the spectra, wavelength calibrates,
flatfields the data and forms the data cube.  
These were registered and stacked using the {\sc exp\_combine} routine.
The (much) higher throughput and higher
spectra resolution ($\lambda/\Delta\lambda\!=\!3000$) of {\sl MUSE}
yielded greatly improved 
signal-to-noise in both continuum and emission lines (see Appendix~\ref{sec:vimosmuse} for a comparison).
For all our analysis of the background sources, we therefore use only the {\sl MUSE} data.

\begin{figure}
\begin{center}
\includegraphics[width=0.47\textwidth]{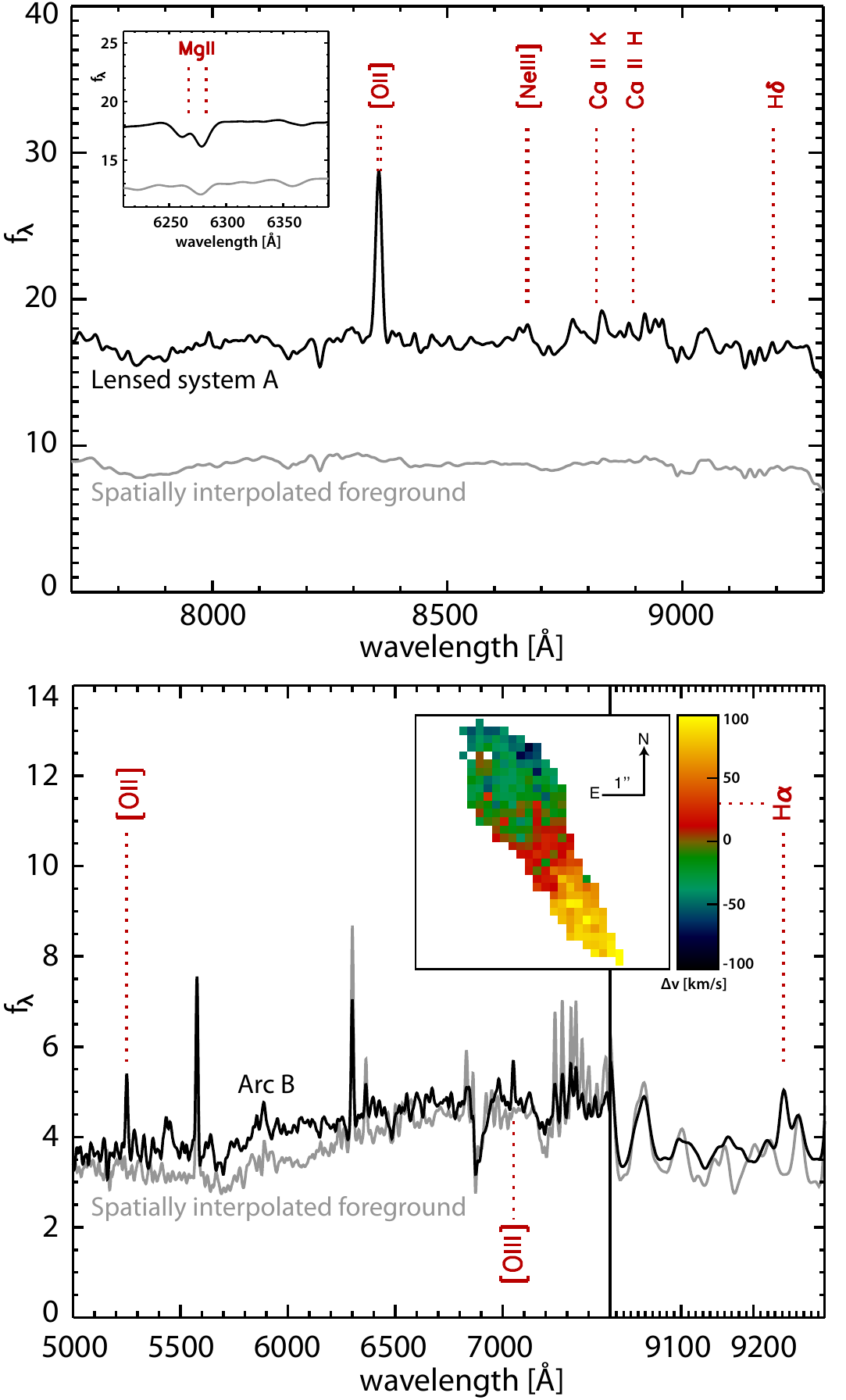}\\
\caption{Observed spectra at the locations of background galaxies A (top panel) and B (bottom panel), smoothed for clarity with a Gaussian of width $5$\,\AA.
In both cases, the grey line shows the spectrum of nearby emission from the foreground cluster, spatially interpolated to the position of the background galaxy.
The coloured insert shows galaxy B's 2D velocity field in a $6\arcsec\!\times6\arcsec$ region, for all IFU pixels where H$\alpha$ emission is detected at signal to noise $>4$. It indicates a rotationally supported disc.}
\label{fig:spectrumA}
\end{center}
\end{figure}

\begin{figure*}
\begin{center}
vspace{-1mm}
\includegraphics[width=0.95\textwidth]{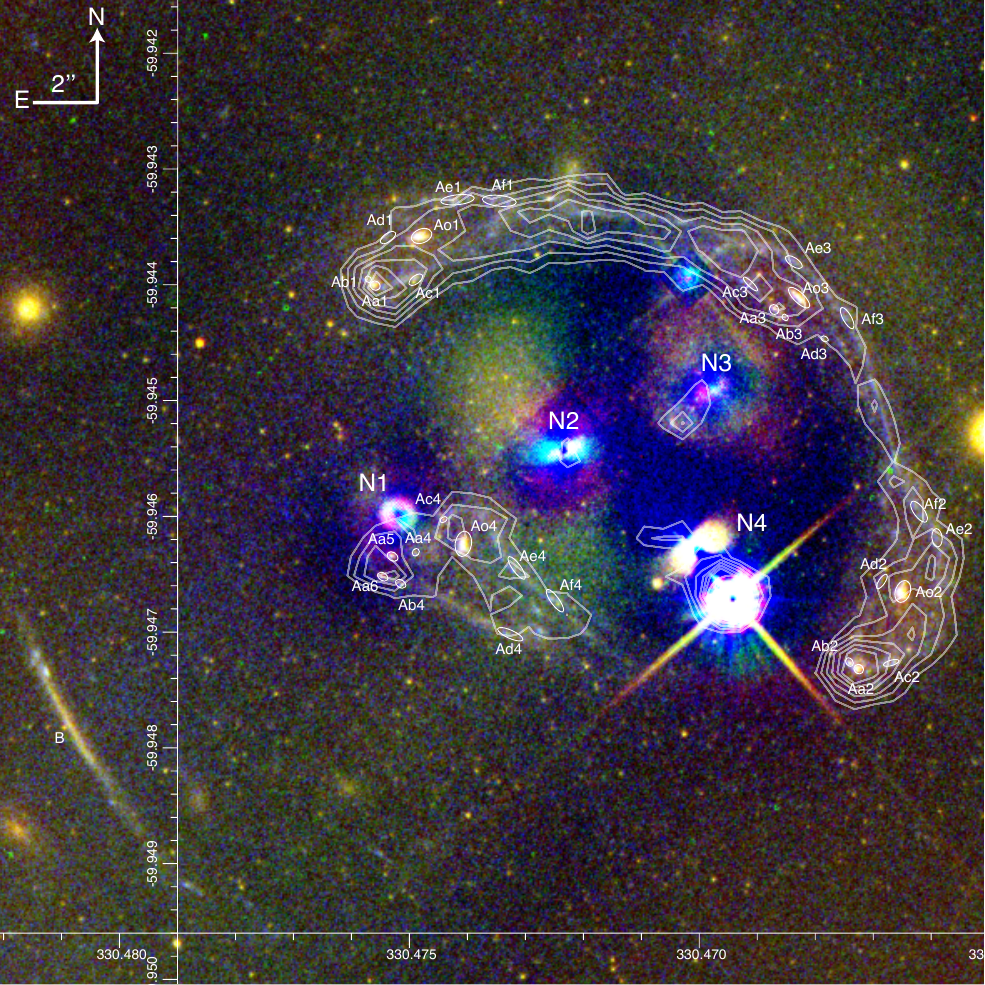}\\
\caption{{\sl Hubble Space Telescope} image of the core of Abell~3827,
  with light from the four bright galaxies subtracted to reveal the
  background lens system.  Colours show the F814W (red), F606W (green)
  and F336W (blue) bands, and the colour scale is square root.
  Linearly spaced contours show line emission at $835.5$\,nm from {\sl
  VLT\,/\,MUSE} integral field spectroscopy. 
  Residual emission near galaxies N.3 and N.4 may be a demagnified fifth image or merely imperfect foreground subtraction, so we do not use it in our analysis.}
\label{fig:imagesub}
\end{center}
\end{figure*}

In lensed system A, the {\sl MUSE} data resolved the 835.5\,nm emission line 
as the [O{\sc ii}]\,$\lambda$3726.8,\,3729.2 doublet at redshift
$z_{\mathrm A}\!=\!1.24145\pm0.00002$, confirmed by the additional
identification of Mg{\sc ii} absorption at 627.0\,nm (figure~\ref{fig:spectrumA}).  
The {\sl MUSE} data also allowed a much improved continuum image to be constructed around
the [O{\sc ii}] emission, then subtracted to leave a higher fidelity [O{\sc ii}]
narrow-band image (figure~\ref{fig:imagesub}).  

For arc B, the {\sl MUSE} data confirms \citet{car10}'s {\sl Gemini} long-slit redshift, finding 
$z_{\mathrm B}\!=\!0.4082\pm0.0001$ at the centre of the arc.
Two blue knots at the north end dominate the [O{\sc ii}] and [O{\sc iii}] emission, but H$\alpha$ emission is visible across its entire length (all three features are at the same redshift).
A 2D map of the best-fit wavelength of the H$\alpha$ emission shows a monotonic velocity gradient of $\sim200$\,km/s from north to south (figure~\ref{fig:spectrumA}).
We are confident that no multiple images are present, having inspected both the {\sl HST} imaging and narrow-band images created from the {\sl MUSE} data at the wavelengths of the emission lines.

\section{Analysis} \label{sec:analysis}

\subsection{Modelling the cluster light distribution}

\begin{table*}
 \centering
  \caption{Total integrated flux of the bright central galaxies, and their derived stellar masses.
  In both {\sl HST}/{\sl ACS} bands (independently), we use {\sc galfit} \citep{galfit} to simultaneously fit the emission from all four galaxies.
  Asterisks denote that a two-component (co-centered) Sersic model was preferred.
  Positions are listed from the F814W band, and are consistent with those from the F606W band.
  Stellar masses $M_*$ interpret the single-band AB magnitude flux via the models of \citet{bc03}, assuming a \citet{cha03} IMF, solar metallicity, and formation redshift  $z_{\mathrm f}\!=\!3$.
  Redshifts, $3\sigma$ upper limits on H$\alpha$ flux (interpreted as limits on star formation rate via \citealt{ken98}, also converted to a \citealt{cha03} IMF), and stellar velocity dispersions are measured from {\sl VLT} spectroscopy.
  \label{tab:stellarmasses}}
  \begin{tabular}{lllr@{$\pm$}llr@{$\times$}llr@{$\times$}lr@{$\times$}lll}
  \hline
  \hline
\multicolumn{5}{l}{~} & \multicolumn{3}{l}{F814W band} & \multicolumn{3}{l}{F606W band} & \multicolumn{2}{l}{H$\alpha$ flux} & SFR & $\sigma_\mathrm{v}^*$ \\
~ & RA & Dec & \multicolumn{2}{l}{$z$} & mag & \multicolumn{2}{l}{$M^*$ [$M_\odot$]} & mag & \multicolumn{2}{l}{$M^*$ [$M_\odot$]} & \multicolumn{2}{l}{[erg/s/cm$^2$]} & [$M_\odot/\!$yr] & [km/s]\!\! \\
\hline
\!N.1 & $330.47518$ & $-59.945997$ & 0.09891 & 0.00032 & $16.86$ & $1.04$ & $10^{11}$ & $17.52$ & $1.00$ & $10^{11}$ & $<6.04$ & $10^{-16}$ & $<0.14$ & 332 \\
\!N.2 & $330.47233$ & $-59.945439$ & 0.09928 & 0.00017 & $15.74^*\!$ & $2.92$ & $10^{11}$ & $16.55$ & $2.46$ & $10^{11}$ & $<2.59$ & $10^{-16}$ & $<0.06$ & 377 \\
\!N.3 & $330.46978$ & $-59.944903$ & 0.09973 & 0.00016 & $15.69^*\!$ & $3.06$ & $10^{11}$ & $16.42^*\!$ & $2.77$ & $10^{11}$ & $<6.31$ & $10^{-16}$ & $<0.14$ & 326 \\
\!N.4 & $330.46999$ & $-59.946322$ & 0.09636 & 0.00026 & $16.18$ & $1.94$ & $10^{11}$ & $16.73$ & $2.08$ & $10^{11}$ & $<1.57$ & $10^{-15}$ & $<0.26$ & 192 \\
\hline
\hline
  \end{tabular}
\end{table*}

It is apparent from the high resolution imaging
(figure~\ref{fig:image}) and our IFU spectroscopy that Abell~3827
contains four bright central galaxies, N.1--N.4.  The object labelled N.5
by \citet{car10} is a Milky Way star: it is a point source in the {\sl
  HST} imaging, and its spectrum contains $z\!=\!0$ {Ca\,{\sc ii}} H
and K absorption lines that are not present in adjacent sources.
Their spectroscopy of N.5 was probably contaminated by the nearby
bright cluster galaxies and diffuse intra-cluster light.  On the other
hand, the westernmost object that \citet{car10} identified as a star,
is actually a faint cluster member galaxy at $z=0.1000\pm0.0002$.  
To avoid confusion, we denote this galaxy N.6.

In the optical {\sl HST} imaging, we use {\sc galfit} \citep{galfit}
to simultaneously fit the light distribution from the four bright
galaxies and the two stars. Most galaxies are well-fit by a single
component model with a Sersic profile, although a double-component
model using two Sersic profiles (with the same centre) is preferred
for galaxy N.3 (and galaxy N.2 in the F814W band). Positions in the
F814W band are listed in table~\ref{tab:stellarmasses}; those in F606W
are consistent within $0\farcs006$ for N.1--3 and, $0\farcs061$
for N.4, due to its proximity to a bright star. To model emission
from the stars, we shift and rescale an isolated star in the same
image. The best-fit galaxy fluxes do not depend significantly upon
the choice of isolated star or the alternative use of a {\sc TinyTim}
model star \citep{tinytim,rho07}.
The photometric errors are dominated by our assumption of analytic functions to fit the light profiles.
The fluxes are likely to be an upper limit because they are computed by integrating these analytic functions to infinite radius, and may also include a component of diffuse intra-cluster light.

In the {\sl VLT}/{\sl VIMOS} spectroscopy, we measure the redshift of galaxies
N.1--N.4 and N.6 by fitting a Gaussian to the H, K and G-band
absorption features.  None of the cluster galaxies exhibits H$\alpha$
line emission, although we attempt to fit a Gaussian at its redshifted
wavelength to obtain $3\sigma$ upper limits on the H$\alpha$ flux.  To
measure the stellar velocity dispersion, we cross-correlate our
spectra with broadened stellar templates from \cite{vaz99}.  These
measurements are presented in table~\ref{tab:stellarmasses}.

\subsection{Strong lens identifications} \label{sec:slids}

\begin{table}
 \centering
  \caption{Locations of multiply imaged systems. 
  Images Ao.$n$ are the bulge, and images A[a--f].$n$ are knots of star formation in the spiral arms.
  Index $n$ is sorted in order of arrival time according to our fiducial model (see table~\ref{tab:pots}).
  Columns denote the ID and position of the image, its major and minor axes, and the angle of its major axis on the sky, anticlockwise from west.
  \label{tab:arcs}}
  \begin{tabular}{lllllr}
  \hline
  \hline
Name & RA & Dec &  \!\!Major & \!\!Minor & \!\!\!Angle \\
\hline
Ao.1 & $330.47479$ & $-59.943580$ & $0\farcs33$ & $0\farcs22$ & $25^\circ$ \\
Ao.2 & $330.46649$ & $-59.946650$ & $0\farcs35$ & $0\farcs23$ & $75^\circ$ \\
Ao.3 & $330.46828$ & $-59.944112$ & $0\farcs43$ & $0\farcs16$ & $140^\circ$ \\
Ao.4 & $330.47407$ & $-59.946239$ & $0\farcs39$ & $0\farcs25$ & $85^\circ$ \\
Aa.1 & $330.47559$ & $-59.944009$ & $0\farcs16$ & $0\farcs14$ & $151^\circ$ \\
Aa.2 & $330.46725$ & $-59.947321$ & $0\farcs16$ & $0\farcs14$ & $140^\circ$ \\
Aa.3 & $330.46871$ & $-59.944215$ & $0\farcs16$ & $0\farcs14$ & $131^\circ$ \\
Aa.4 & $330.47489$ & $-59.946312$ & $0\farcs12$ & $0\farcs10$ & $54^\circ$ \\
Aa.5 & $330.47529$ & $-59.946349$ & $0\farcs18$ & $0\farcs13$ & $150^\circ$ \\
Aa.6 & $330.47546$ & $-59.946523$ & $0\farcs18$ & $0\farcs12$ & $150^\circ$ \\
Ab.1 & $330.47571$ & $-59.943954$ & $0\farcs11$ & $0\farcs09$ & $131^\circ$ \\
Ab.2 & $330.46741$ & $-59.947260$ & $0\farcs14$ & $0\farcs11$ & $131^\circ$ \\
Ab.3 & $330.46852$ & $-59.944283$ & $0\farcs11$ & $0\farcs09$ & $131^\circ$ \\
Ab.4 & $330.47515$ & $-59.946584$ & $0\farcs18$ & $0\farcs12$ & $150^\circ$ \\
Ac.1 & $330.47489$ & $-59.943958$ & $0\farcs25$ & $0\farcs13$ & $41^\circ$ \\
Ac.2 & $330.46669$ & $-59.947267$ & $0\farcs25$ & $0\farcs08$ & $20^\circ$ \\
Ac.3 & $330.46912$ & $-59.943994$ & $0\farcs30$ & $0\farcs08$ & $140^\circ$ \\
Ac.4 & $330.47441$ & $-59.946030$ & $0\farcs12$ & $0\farcs08$ & $220^\circ$ \\
Ad.1 & $330.47537$ & $-59.943594$ & $0\farcs28$ & $0\farcs13$ & $40^\circ$ \\
Ad.2 & $330.46685$ & $-59.946564$ & $0\farcs26$ & $0\farcs10$ & $60^\circ$ \\
Ad.3 & $330.46784$ & $-59.944468$ & $0\farcs12$ & $0\farcs08$ & $157^\circ$ \\
Ad.4 & $330.47326$ & $-59.947020$ & $0\farcs42$ & $0\farcs13$ & $160^\circ$ \\
Ae.1 & $330.47345$ & $-59.943276$ & $0\farcs53$ & $0\farcs17$ & $178^\circ$ \\
Ae.2 & $330.46590$ & $-59.946186$ & $0\farcs28$ & $0\farcs16$ & $100^\circ$ \\
Ae.3 & $330.46837$ & $-59.943805$ & $0\farcs30$ & $0\farcs13$ & $150^\circ$ \\
Ae.4 & $330.47315$ & $-59.946447$ & $0\farcs42$ & $0\farcs13$ & $130^\circ$ \\
Af.1 & $330.47417$ & $-59.943267$ & $0\farcs52$ & $0\farcs15$ & $10^\circ$ \\
Af.2 & $330.46621$ & $-59.945961$ & $0\farcs39$ & $0\farcs16$ & $130^\circ$ \\
Af.3 & $330.46745$ & $-59.944289$ & $0\farcs37$ & $0\farcs13$ & $123^\circ$ \\
Af.4 & $330.47249$ & $-59.946730$ & $0\farcs42$ & $0\farcs13$ & $130^\circ$ \\ 
\hline
\hline
  \end{tabular}
\end{table}

Figure~\ref{fig:imagesub} presents a multicolour image of Abell~3827,
after subtracting the best-fit model of optical emission from the
optical {\sl HST} bands to reveal the morphology of the gravitationally lensed system.
In the UV {\sl HST} imaging, the contrast between cluster member
galaxies and the background lens system is much lower, so we do not
fit and subtract the foreground flux.

Lensed image A is an almost face-on spiral galaxy, with a bulge and
many resolved knots of star formation that can all be used as
independent lensed sources.  The association of knots between multiple
images is not perfectly clear, due to the bright intra-cluster light
and a surprising density of point sources, particularly near galaxy
N.1; we present the most likely identifications in
table~\ref{tab:arcs}, but analyse alternative configurations in
appendix~\ref{sec:altconfigs}.
Contours in figure~\ref{fig:imagesub} show a (continuum subtracted) narrow-band image created
by subtracting continuum emission from each spatial pixel (fitted
using a low-order polynomial over the wavelength range
$830$--$840$\,nm) then collapsing the {\sl MUSE} IFU data cube over
$\pm$300\,km\,s$^{-1}$ from the peak of the emission.  
The 2D map of this line emission matches precisely the lensed galaxy's broad-band morphology.
Variations in the intensity of the line emission can be explained by lensing magnification. 
According to our fiducial {\sc Lenstool} model (see section~\ref{sec:resultsLenstool}), the magnification at the position of the bulge images, and variation across the spiral is 
$1.29^{+0.10}_{-0.15}$ for image 1,
$0.95^{+0.06}_{-0.03}$ for images 2 and 3, but
$1.62^{+0.43}_{-0.28}$ for image 4 (with the magnification greatest near galaxy N.1).

The 835.5\,nm emission near the galaxies N.4 and N.3 is possibly a demagnified image of bulge Ao and knot Aa.
A demagnified image of the bulge is robustly predicted between N.2 and N.4, although our fiducial {\sc Lenstool} model places it closer to N.2 at (330.47135,\,-59.945850).
Models allowing a demagnified image of Aa stretched towards N.3 are possible, but at lower likelihood, and (for reasonable positions of the cluster-scale halo) this would be {\em in addition} to a demagnified image between N.2 and N.4.
Since these identifications are not robust with current data, and could be merely imperfect foreground subtraction near the bright cluster galaxies, we exclude them from our strong lensing analysis

Arc B would be difficult to interpret as a strong lens, as previously suggested, because it is at lower redshift than system A but greater projected distance from the lens (whereas Einstein radius should increase with redshift).
Instead, its constant velocity gradient suggests that is merely an edge-on spiral galaxy, aligned by chance at a tangential angle to the cluster, elongated and flexed (our fiducial {\sc Lenstool} model predicts shear $\gamma=0.20\pm0.01$) but only singly imaged. 
The two knots of star formation at its north end further enhance its visual appearance of curvature.
This is consistent with the absence of observed counterpart images, and the lack of fold structure in its 2D velocity field.
We therefore exclude arc B from our strong lensing analysis.

\subsection{Modelling the cluster mass distribution}

To model the strong gravitational lens system, we use two independent
software packages: {\sc Grale} \citep{lie06} and {\sc Lenstool}
\citep{Lenstool}.  The two packages make very different assumptions.
      {\sc Grale} models the mass distribution using a free-form grid,
      in which the projected density at each pixel is individually
      constrained and individually adjustable; {\sc Lenstool}
      interpolates a parametric model built from a relatively small
      number of components, each of which has a shape that matches the
      typical shapes of clusters.  The two packages also exploit
      slightly different features of the input data.  For example,
      both methods match the position of multiply-imaged systems, but
      {\sc Grale} can also match their shape, and use the absence of
      counter images where none are observed; while {\sc Lenstool} can
      expoit the symmetries of great arcs.

\subsubsection{Grale}\label{sec:methodGrale}

{\sc Grale} is a free-form, adaptive grid method that uses a genetic algorithm to iteratively refine the
mass map solution \citep{lie06,lie07,lie08a,lie08b,lie09,lie12}. 
We work within a $50\arcsec\times50\arcsec$ reconstruction region, centered on ($330.47043,\,-59.945804$).
An initial coarse resolution grid is populated with a basis set; in this work we use projected Plummer
density profiles \citep{plum11,dejong87}. A uniform mass sheet covering the whole modelling region can also be added to supplement the basis set.  As the code runs, the more dense regions are resolved with a finer grid, with each
cell having a Plummer with a proportionate width. 
The code is started with an initial set of trial
solutions. These solutions, as well as all the later evolved ones are evaluated for
genetic fitness, and the fit ones are cloned, combined and mutated. 
The resolution is determined by the number of Plummers used. The initial coarse
resolution grid is refined nine times, allowing for more detail in the
reconstruction; the best map is selected based on the fitness measure.
The final map consists of a superposition of a mass sheet and many Plummers,
typically a few hundred to a thousand, each with its own size and weight, determined by the genetic algorithm.
Note that adopting a specific (Plummer) density profile for our basis set does not at all restrict 
the profile shapes of the mass clumps in the mass maps.

We use two types of fitness measures in this work. 
(a)~{\it Image positions.}  A successful mass map would lens image-plane images of the same source back to
the same source location and shape. 
We take into account the position, shape, and angular extent of all the images in table~\ref{tab:arcs}, by
representing each image as a collection of points that define an area. A mass map has a greater fitness
measure if the images have a greater fractional area overlap in the source plane. This ensures against
over-focusing, or over-magnifying images, which plagued some of the early lens reconstruction methods.
(b)~{\it Null space.} Regions of the image plane that definitely do not contain any lensed features belong to
the null space. Each source has its own null space. Typically, a null space is all of the image plane, with `holes' for the observed images, and suspected counter images, if any. The product of these two fitness measures is used to select the best map in each reconstruction.

All lensed images in this cluster arise from extended sources (star formation knots within a galaxy). 
Because of that it is hard to identify the center of each image to a precision comparable to HST resolution. 
The $0\farcs3$--$0\farcs6$ extent of {\sc Grale} points representing some image will be, approximately, the lower bound on the lens plane rms between the observed and predicted images.

We run twenty mass reconstructions for each image configuration (a limit set by computational time constraints), and the maps that are presented here are the averages of these.
The full range of the recovered maps provides an estimate of the statistical error in the mass maps. 

\subsubsection{Lenstool}

{\sc Lenstool} is a parametric method that uses a Markov-Chain Monte Carlo (MCMC) fit with a comparatively smaller number of mass peaks, but which are free to move and change shape.
We construct our mass model using one dual Pseudo Isothermal Elliptical \citep[dPIE,][]{limousin05,eliasdottir07} halo for the overall cluster, plus a smaller Pseudo Isothermal Elliptical halo for each galaxy N.1--4 and N.6.
Each halo is characterised by a position $(x,\,y)$, velocity dispersion $\sigma_\mathrm{v}$, ellipticity $e$, and truncation radius $r_{\rm cut}$; the cluster halo is also allowed to have a nonzero core radius $r_{\rm core}$.
We set the following priors on the cluster halo: $e\!<\!0.75$, $r_{\rm core}\!<\!4\arcsec$, and the position has Gaussian probability with width $\sigma\!=\!2\arcsec$ centered on N.2.
For the galaxy halos, we set a prior $e\!<\!0.45$.
The position of N.1 is of particular scientific interest, and will be well-constrained because it is surrounded by strong lens images, so to avoid any bias, we set a prior that is flat within $-5\arcsec\!<x<3\arcsec$ and $-3\arcsec\!<y<3\arcsec$ of the optical emission.
The position of N.2--N.4 will be less well constrained, so we set Gaussian priors\footnote{We check that the Gaussian priors do not bias the best-fitting position by shifting the priors to the peaks in the posterior and rerunning the analysis. Positions all move by less than $0\farcs08$ in $x$ and $y$, indicating convergence to within statistical errors.} with width $\sigma\!=\!0\farcs5$, centred on their optical emission. 
The parameters of N.6 are poorly constrained, because it is faint and far from the strong lens systems, so we fix its position to that of its optical emission, and fix $e\!=\!0$ to reduce the search dimensionality.
The strong lensing data alone provide no constraints on the outer regions of the mass distribution, so we manually fix $r_{\rm cut}=1000\arcsec$ for the cluster halo, and $r_{\rm cut}=100\arcsec$ for the galaxy halos, well outside any region of influence.

As constraints, we use the positions of all multiple images in table~\ref{tab:arcs} and, following additional symmetries of the image, require the $z_{\mathrm A}\!=\!1.24$ critical curve to pass through $(330.47113,-59.943529)$ and $(330.46677,-59.945195)$ perpendicular to an angle of $175^\circ$ and $110^\circ$ respectively.
These are indicated in the bottom panel of figure~\ref{fig:configAGrale}.
We then optimise the model in the image plane using an MCMC search with parameter {\sc BayesRate}$=0.1$ \citep[which allows the posterior to be well explored during burn-in, before converging to the best-fit solution; for more details, see][]{kneib96,smith05,jauzac14a}.
We assume an error of $0\farcs2$ (68\% CL) on every position. 
This choice merely rescales the posterior.
If we instead assume an error of $0\farcs273$, the model achieves reduced $\chi^2/\mathrm{dof}\!=\!1$.

\section{Results} \label{sec:results}

\subsection{Stellar mass}

Spectroscopic redshifts of N.1--N.4 and N.6 confirm that they are all $z_{c\ell}\!\approx\!0.099$ cluster members.
Notably, N.1--N.3 are at essentially the same redshift as each other (which is consistent with the mean redshift of all known member galaxies), and are projected along a straight line.
One the other hand, N.4 has a relative line-of-sight velocity $\sim\!900\,$km/s.

In table~\ref{tab:stellarmasses}, we interpret the galaxies' integrated broad-band fluxes (in individual filters) as stellar masses $M^*$ via the models of \citet{bc03}, assuming a \citet{cha03} initial mass function (IMF), solar metallicity, and formation redshift $z_{\mathrm f}\!=\!3$.
The inferred stellar mass of N.1 is surprisingly low compared to its stellar velocity dispersion, but this could be because it is furthest from the cluster core, so its measured flux contains the least contamination from intra-cluster light.

We also interpret the $3\sigma$ upper limits on the H$\alpha$ flux as star formation rate (SFR), following \cite{ken98} relations converted to a \cite{cha03} IMF. 
The red and dead ellipticals exhibit effectively zero star formation, consistent with their faint UV broad-band fluxes.

\subsection{Total mass distribution} \label{sec:resultsdm}

\subsubsection{Grale} \label{sec:resultsGrale}

\begin{figure}
\begin{center}
\includegraphics[trim = 8mm 72mm 8mm 37mm, clip, width=0.418\textwidth]{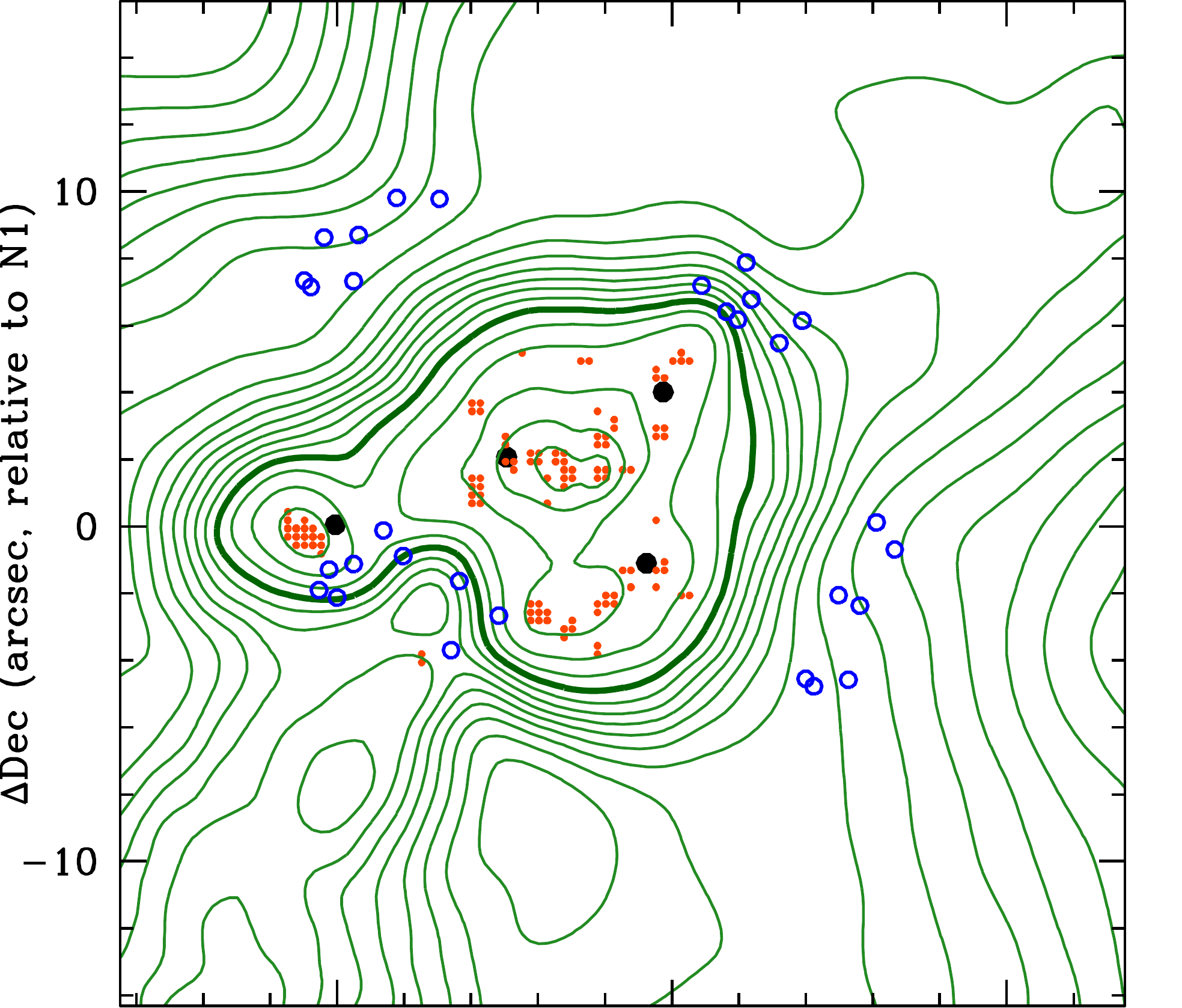}\\
\includegraphics[trim = 8mm 72mm 8mm 37mm, clip, width=0.418\textwidth]{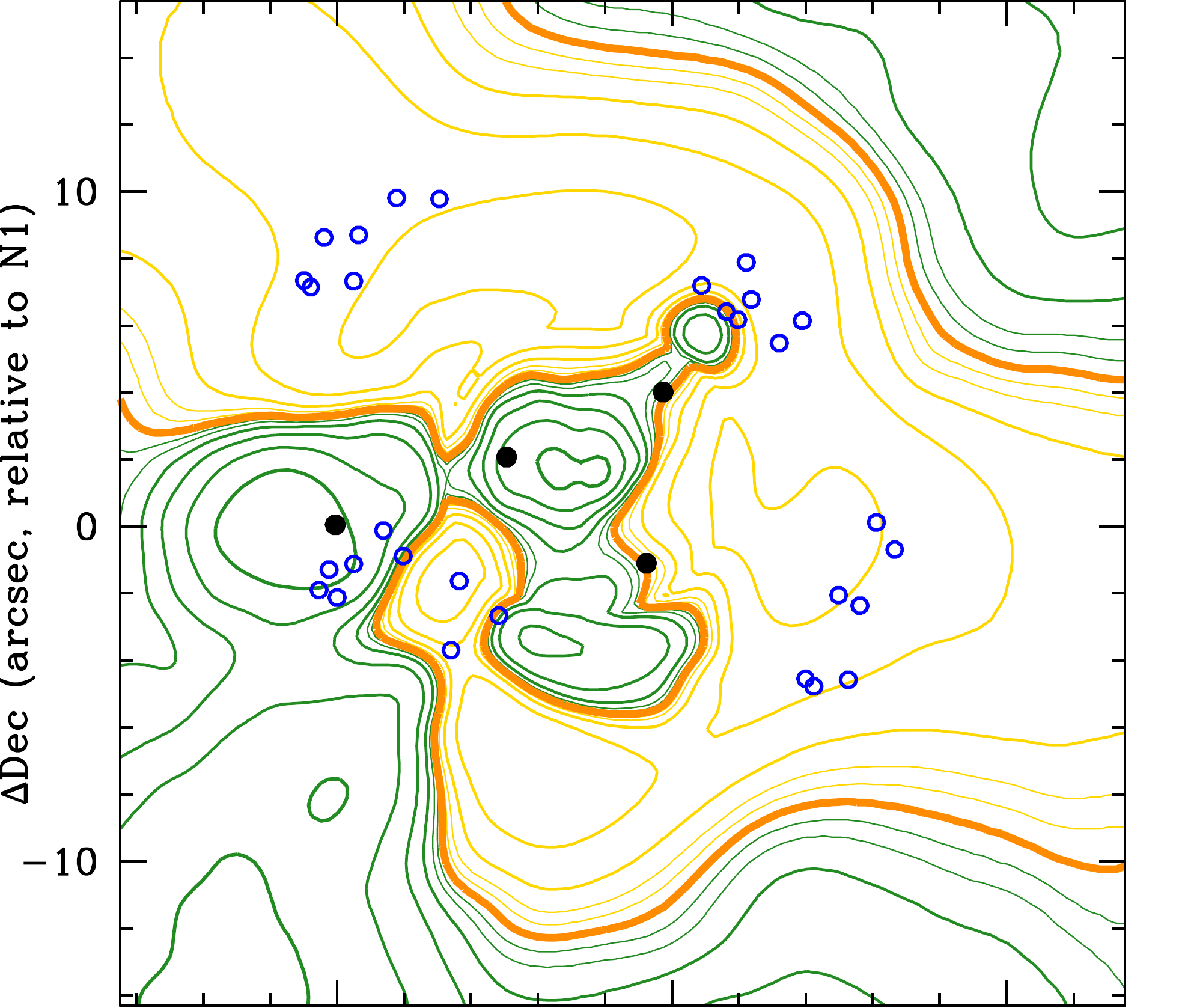}\\
\includegraphics[trim = 8mm 54mm 8mm 37mm, clip, width=0.418\textwidth]{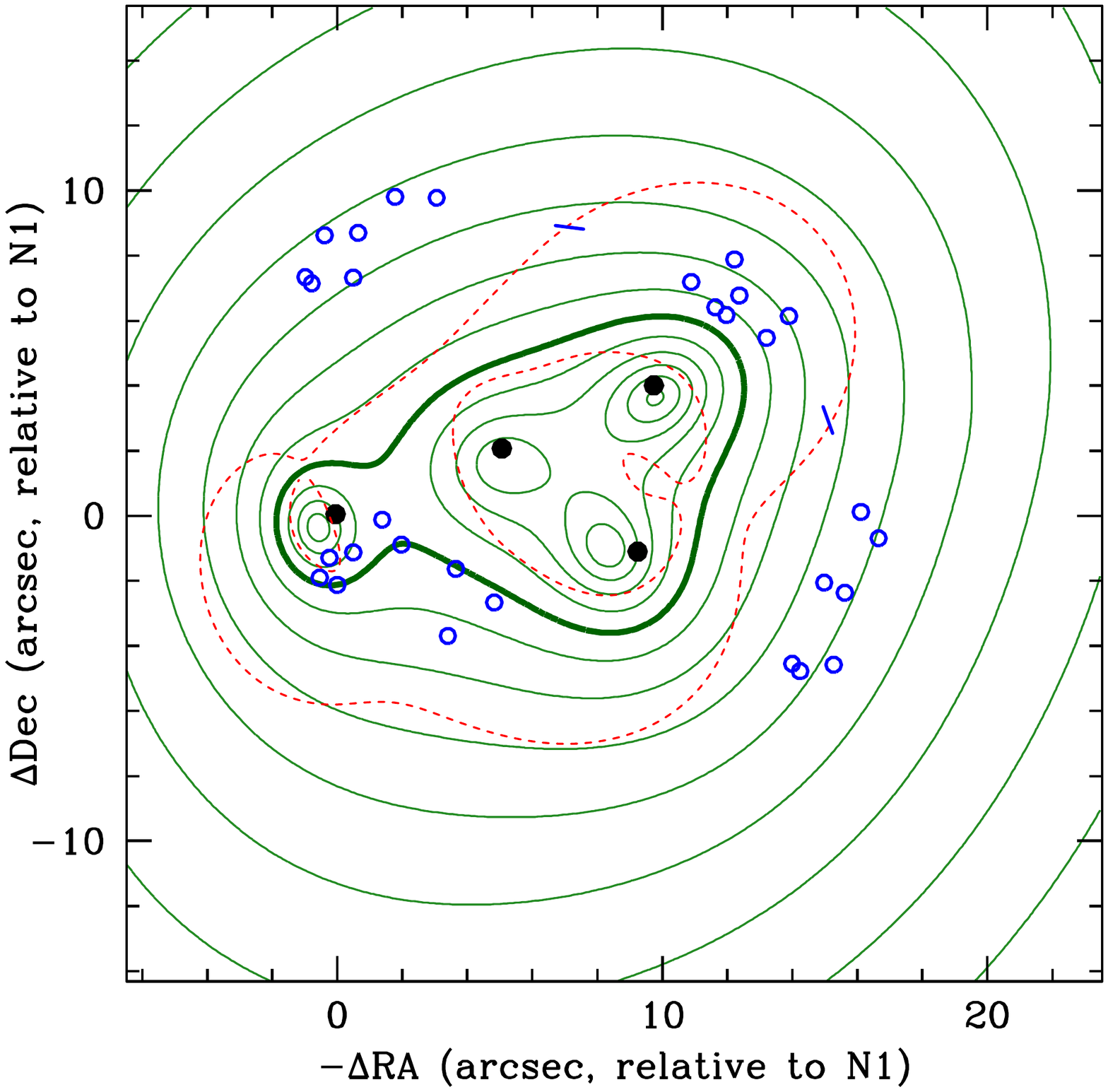}
\caption{
{\it Top panel:} Map of total mass in the cluster core, reconstructed using {\sc Grale}.
Green contours show the projected mass density, spaced logarithmically by a factor 1.15; the thick contour shows convergence $\kappa\!=\!1$ for $z_{c\ell}\!=\!0.099$ and $z_\mathrm{A}\!=\!1.24$ ($\Sigma_\mathrm{crit}\!=\!1.03\,$g/cm$^2$). 
Red dots show local maxima in individual realisations of the mass map. 
Black dots show cluster ellipticals N.1--N.4. 
Blue circles show the lensed images.
{\it Middle panel:} Mass after subtracting a smooth cluster-scale halo to highlight substructure.
The thick contour is at $\Delta\kappa\!=\!0$.
The green (positive) and yellow (negative) contours are at $\Delta\kappa\!=\!\pm0.025,\pm0.05,\pm0.1,\pm0.2$\ldots.
{\it Bottom panel:} Total mass, as in the top panel but reconstructed via {\sc Lenstool}.
The red dashes show the $z_\mathrm{A}\!=\!1.24$ critical curve.}
\label{fig:configAGrale}
\end{center}
\end{figure}

The {\sc Grale} reconstruction of the projected mass distribution within the central $30\arcsec\!\times30\arcsec$ of the $50\arcsec\!\times50\arcsec$ fitted region is shown in the top panel of figure~\ref{fig:configAGrale}.
The mean rms offset of observed lens images from their predicted positions in the image plane is $\langle\mathrm{rms}_i\rangle\!=\!0\farcs34$, but this value is inflated by the size of the input images, as explained in Section~\ref{sec:methodGrale}.

The mass distribution peaks at (330.47068,\,-59.945513), i.e.\ ($8\farcs15$,\,$1\farcs73$) or $r\!=\!8\farcs33$ from galaxy N.1.
The total mass of the cluster core inside a cylinder with this radius and centered on the mass peak is $M_{c\ell}\!=\!3.58\times10^{12}\,M_\odot$.
The typical fractional error in surface mass density in the central region is $\sim\!15$\%.
The total mass projected within $1\farcs5\!=\!2.7$\,kpc of galaxies N.1--4 is $1.63$,  $2.02$,  $1.70$ and $1.67\times10^{11}\,M_\odot$ respectively. 

\begin{table*}
 \centering
  \caption{Parameters of the best-fit, fiducial mass model constructed by {\sc Lenstool}. 
  Positions are relative to the peak of light emission, except for the cluster-scale halo, whose position is relative to the peak of emission from galaxy N.1.
  Quantities in square brackets are not fitted.
  Errors on other quantities show 68\% statistical confidence limits due to uncertainty in the lensed image positions, marginalising over uncertainty in all other parameters.
  \label{tab:pots}}
  \begin{tabular}{lr@{}lr@{}lr@{}lr@{}lr@{}lr@{}lccc}
  \hline
  \hline
~ & \multicolumn{2}{l}{$\sigma_\mathrm{v}\,$[km/s]\!\!\!} & \multicolumn{2}{c}{$x\,$[\arcsec]} & \multicolumn{2}{c}{$y\,$[\arcsec]} & \multicolumn{2}{c}{$e$} & \multicolumn{2}{c}{$\theta\,$[$^\circ$]} & \multicolumn{2}{c}{$r_\mathrm{core}\,$[\arcsec]}  
& $\langle\mathrm{rms}_i\rangle\,$[\arcsec] & $\chi^2\!/\mathrm{dof}$ & $\log_{10}{\!(E)}$ \\
\hline 
\multicolumn{13}{l}{Fiducial model: } & $0.26$ & $49.3/23$ & $-26.4$ \\
N.1 & $190$ & $^{+8}_{-12}$ & $-0.61$ & $^{+0.14}_{-0.12}$ & $-0.46$ & $^{+0.20}_{-0.14}$ & $0.25$ & $^{+0.15}_{-0.04}$ & $101$ & $^{+22}_{-22}$ & \multicolumn{2}{c}{[$\to$0]} \\ 
N.2 & $219$ & $^{+18}_{-38}$ & $-0.13$ & $^{+0.28}_{-0.46}$ & $-0.48$ & $^{+0.30}_{-0.30}$ & $0.09$ & $^{+0.12}_{-0.09}$ & $174$ & $^{+22}_{-37}$ & \multicolumn{2}{c}{[$\to$0]} \\ 
N.3 & $254$ & $^{+17}_{-14}$ & $0.09$ & $^{+0.25}_{-0.25}$ & $-0.36$ & $^{+0.18}_{-0.29}$ & $0.25$ & $^{+0.04}_{-0.10}$ & $30$ & $^{+11}_{-13}$ & \multicolumn{2}{c}{[$\to$0]} \\ 
N.4 & $235$ & $^{+20}_{-34}$ & $-0.99$ & $^{+0.39}_{-0.34}$ & $-0.01$ & $^{+0.35}_{-0.27}$ & $0.19$ & $^{+0.12}_{-0.09}$ & $121$ & $^{+22}_{-54}$ & \multicolumn{2}{c}{[$\to$0]} \\ 
N.6 & $18$ & $^{+44}_{-1}$ & \multicolumn{2}{c}{[0]} & \multicolumn{2}{c}{[0]} & \multicolumn{2}{c}{[0]} & \multicolumn{2}{c}{[0]} & \multicolumn{2}{c}{[$\to$0]} \\ 
Cluster & $620$ & $^{+101}_{-58}$ & $6.18$ & $^{+1.33}_{-1.04}$ & $2.30$ & $^{+1.86}_{-1.51}$ & $0.70$ & $^{+0.01}_{-0.24}$ & $61$ & $^{+3}_{-4}$ & $30.12$ & $^{+9.23}_{-6.43}$ \\ 
\hline\hline
  \end{tabular}
\end{table*}

To highlight the {\em position} of substructures, the red dots in the top panel of figure~\ref{fig:configAGrale} show the local maxima in each of the individual, statistically independent realisations of the mass map (within the central $\kappa\!=\!1$ contour only), and the middle panel of figure~\ref{fig:configAGrale} shows the mean map after subtracting a smooth, cluster-scale halo.
The cluster-scale halo used is centered on the peak density, has constant projected density inside a core of radius $4\farcs4\!=\!8.0$\,kpc, and a projected density profile $\rho_\mathrm{2D}(r)\propto r^{-1.3}$ outside this core.

The most robustly constrained region of interest is near galaxy N.1, owing to the proximity of several lensed images.
Local peaks in individual realisations of the mass map form a tight cluster $1\farcs01\pm0.39$ east-southeast of the peak of optical emission, implying that the offset is significant at the $2.6\sigma$ level.
The mass within $1\farcs5$ of this location is $1.66\times10^{11}\,M_\odot$.

Galaxy N.3 is next nearest to gravitationally lensed images. 
The closest local mass peak is frequently $1$--$2\arcsec$ northwest of the optical emission, and the mass within $1.5\arcsec$ of this location is $1.38\times10^{11}\,M_\odot$.
Statistically significant structure is apparent in the mean mass map, but its presence is less robust than for N.1.
It may be interesting that galaxies N.1--3, the mass peaks closest to N.1 and N.3, and the cluster's large-scale diffuse light all lie close to a straight line. 
Infall from preferred directions along filaments is expected \citep[\eg][]{eagle} --- although orbits of galaxies do not generally stay radial this far from the virial radius, so it may also be coincidence.
Galaxy N.4 has no local mass peak in many realisations of the reconstruction; the closest set of peaks is offset $\sim3\arcsec$ to the southeast. 
However, the region around N.4 is not as well constrained by nearby lensed images, and appendix~\ref{sec:altconfigs} shows that the distribution of surrounding mass is sensitive to the assumed identification of strong lens images.
The position of the mass associated with galaxy N.2 is difficult to disentangle from that of the cluster-scale halo.

\subsubsection{Lenstool} \label{sec:resultsLenstool}

Best-fit parameters for the fiducial mass model are listed in table~\ref{tab:pots}, and the corresponding mass map is shown in the bottom panel of figure~\ref{fig:configAGrale} (it is uninformative to subtract the cluster scale halo, as its core is flatter than the cuspy galaxies, so does not visually affect their position).
The mean rms offset of observed lens images from their predicted positions in the image plane is $\langle\mathrm{rms}_i\rangle\!=\!0\farcs26$ (41\% of which is contributed by systems Ae and Af, whose observed position in some multiple images is indeed the least certain).
With the assumed $0\farcs2$ errors on observed positions, the model achieves $\chi^2\!/\mathrm{dof}\!=\!49.3/23$, and Bayesian evidence $\log_{10}{\!(E)}\!=\!-26.4$.
Some of the {\sc Lenstool} MCMC samples predict fifth and sixth images of knot Ac to the east of Aa.5 and Aa.6, which are indeed faintly visible as the continuation of image A's ring.

The total mass projected within $r\!\!=\!\!8\farcs33$ of ($330.47068$, $-59.945513$) is $M_{c\ell}=3.49\times10^{12}\,M_\odot$.
The total mass within $1\farcs5$ of galaxies N.1--4 is $0.65$,  $1.20$,  $1.48$ and $1.83\times10^{11}\,M_\odot$ respectively.
These numbers include a contribution from the cluster-scale halo, and are most useful for comparison to the {\sc Grale} results.
Integrating the individual components of the mass model analytically within a $1.5\arcsec$ radius, {\it excluding} the cluster halo, yields masses of $0.72$, $0.95$, $1.28$ and $1.10\times10^{11}\,M_\odot$ \citep[see equation~9 of][]{limousin05}.
The mass associated with N.1 increases slightly in this calculation because these measurements are centered on the mass peaks.
The 68\% confidence limits on all of these masses, obtained by propagating the statistical uncertainty (only) in the observed image positions through the MCMC sampler, are approximately 1\%.

The mass associated with galaxy N.1 is offset by $0\farcs76^{+0.13}_{-0.16}$ from the peak of its light emission\footnote{This remains as $0\farcs81^{+0.13}_{-0.12}$ even if $10\%$ of the mass (accounting for the stellar component) is forced to lie on the galaxy; the offset is necessary to ensure the observed multiplicity of Aa.5 and Aa.6.}.
The 68\% confidence limit, again obtained from the MCMC samples, includes only statistical uncertainty propagated from that in the observed image positions.
To estimate the additional uncertainty (bias) caused by the relative inflexibility of {\sc Lenstool}'s parametric model to represent a complex mass distribution, we retry the optimisation using two alternative model configurations.

Refitting the data using a model with a second cluster-scale halo achieves $\langle\mathrm{rms}_i\rangle\!=\!0\farcs25$, although with worse $\chi^2\!/\mathrm{dof}=48.5/17$, and $\log_{10}{\!(E)}\!=\!-28.0$.
In this model, the first cluster halo remains between N.2 and N.3, and the second cluster halo appears between N.3 and N.4, whose masses are reduced by $\sim\!45\%$. 
The mass associated with galaxy N.1 moves to $(-0\farcs68^{+0.13}_{-0.12},\,-0\farcs25^{+0.19}_{-0.12})$.
This $0\farcs22$ shift relative to the fiducial reconstruction provides one estimate of model-induced bias.

Second, we can reuse the fiducial model to perform an MCMC fit to a mass distribution that has a similar configuration to the real distribution but is not impeded by the parametric limitations of {\sc Lenstool}.
We generate four, slightly different realisations of mock data by raytracing the positions of each lensed image (\eg\ A[a-f].1) through the fiducial model to the source plane, then back to (multiple locations in) the image plane. 
In this mock data, the true position of the mass is known, and can be represented perfectly by {\sc Lenstool}.
We perform four independent fits, centering the priors around the true positions of the mass.
The mean spurious offset of N.1 is $0\farcs55\pm0.11$, and the mean spurious offsets of N.2--4 are $0\farcs55\pm0.09$.
Comparing this to the total $\sim\!0\farcs76$ offsets in the real fit therefore suggests a similar $\sim\!0\farcs21$ budget for model-induced bias.

The mean offset of N.2--4 (with the observed constraints) is $0\farcs62$, which could be interpreted to imply a characteristic error on N.1 of this order. 
However, these are measured with a prior, and their uncertainty is much greater than that of N.1 because those galaxies happen to be much farther from strongly lensed systems --- and, in the case of N.2, because of degeneracy with the position of the cluster-scale halo.
Contrary to the results from {\sc Grale}, the mass associated with galaxy N.3 is coincident with the position of its light emission within measurement error; the mass within $1\farcs5$ of {\sc Grale}'s offset peak is a lower $1.07\times10^{11}\,M_\odot$.
The mass associated with N.4 is offset at only marginal statistical significance but, intriguingly, the offset is in the same direction as that measured by {\sc Grale}.

\section{Interpretation}\label{sec:discussion}

We have modelled the distribution of mass in the cluster using two independent approaches: free-form {\sc Grale} and parametric {\sc Lenstool}.
The general agreement between methods is remarkable, both in terms of total mass and many details.

The most striking result is that the mass associated with galaxy N.1 is offset from its stars,
$1\farcs01\pm0.39$ east-southeast with {\sc Grale} or $0\farcs76^{+0.34}_{-0.37}$ southeast with {\sc Lenstool} (linearly adding statistical and method-induced errors).
That the measurements are consistent with each other, and resilient to small changes in the strong lens identifications (see appendix~\ref{sec:altconfigs}), supports a robust conclusion that the offset is real.
To combine the analyses we note that, although they 
start with mostly identical input data, uncertainty on their final constraints is dominated by the highly-nonlinear reconstruction procedures, which are independent.
We therefore average the best-fit values with equal weight and add their errors in quadrature, to infer a combined constraint on the offset between mass and light 
\begin{equation}
\delta=0\farcs89^{+0.26}_{-0.27}=1.62^{+0.47}_{-0.49} \,\mathrm{kpc~(68\%~CL)}.
\end{equation}
The strong lens configuration makes galaxy N.1 the best measured of all the cluster members, but both {\sc Grale} and {\sc Lenstool} provide marginal evidence for an offset in galaxy N.4, and {\sc Grale} suggests a similarly unexpected distribution of mass near in N.3.

Interpreting an offset between mass and stars is difficult. 
It could feasibly be caused by different tidal forces or dynamical friction on the different-sized dark matter/stellar haloes; partially stripped gas (or unrelated foreground/background structures) that contributes to the total mass \citep{eck14,roe14a}; or a displacement of the light emission due to recent star formation triggered in stripped gas.
However, to first order, tidal forces do not alter the peak position.
Archival {\sl Chandra} data also show no substructure near this cluster core. 
The effectively zero star formation we observe also suggests that broad band emission should trace stars that existed before the merger.
Instead, \cite{ws11} interpreted the offset in terms of dark matter's self-interaction cross-section $\sigma_{\mathrm{DM}}/m$, using a toy model of interactions equivalent to an optical depth (see their equation~(3)\footnote{Note that the prefactor in equation~(4) of \cite{ws11} should be $6.0\!\times\!10^{3}$ rather than $6.0\!\times\!10^{4}$.} and see also \citealt{mkn11,kah14,har14}).
If stars in the infalling galaxy are subject only to gravity, but its dark matter also feels an effective drag force, after infall time $t_{\mathrm{infall}}$, dark matter lags behind by an offset
\begin{equation}
\delta(t_\mathrm{infall})\sim
\frac{GM_{c\ell}M_{\mathrm{DM}}}{\pi\, s_{\mathrm{DM}}^2 r_{\mathrm{DM}}^2}\,\frac{\sigma_{\mathrm{DM}}}{m}\;t_{\mathrm{infall}}^2,\label{eqn:disp}
\end{equation}
where $M_{c\ell}$ is the mass of the cluster interior to the infalling galaxy,
which has dark matter mass $M_\mathrm{DM}$ and cross-sectional area $\pi s_{\mathrm{DM}}^2$, at clustercentric radius $r_{\mathrm{DM}}$.
Adopting mean masses from our {\sc Grale} and {\sc Lenstool} analyses, $M_{c\ell}\!=\!3.54\!\times\!10^{12}$,
$M_\mathrm{DM}\!=\!1.19\!\times\!10^{11}$, parameter $s_{\mathrm{DM}}\!=\!4\farcs1$ following \cite{ws11}, and $r_{\mathrm{DM}}\!=\!r\!=\!8\farcs3$, then propagating 10\% errors on the masses and $0\farcs5$ errors on the sizes, suggests
\begin{equation}
\sigma/m\sim (1.7\pm0.7)\!\times\!10^{-4}
\left(\frac{t_\mathrm{infall}}{10^{9}\, {\rm yrs}}\right)^{-2} {\rm cm}^2/{\rm g}.
\label{estim}
\end{equation}
The infall time must be less than $10^{10}$ years, the age of the Universe at the cluster redshift.
Given the lack of observed disruption, collinearity (and common redshift) of N.1--3, they are likely to be infalling on first approach from a filament, and moving within the plane of the sky.
Thus $t_\mathrm{infall}\!\simlt\!10^{9}\,$yrs, the approximate cluster crossing time, and assuming this conservative upper bound places a conservative lower bound on $\sigma/m$.
If any component of the motion is along our line of sight, the 3D offset may be larger, so our assumption of motion exactly within the plane of the sky is also conservative.
Using a different set of strong lens image assignments (see appendix~\ref{sec:altconfigs}), we recover the $6$\,kpc offset and correspondingly larger cross-section of \cite{ws11}.
These image assignments are now ruled out by our new IFU spectroscopy, which unambiguously traces the morphology of the lens, even through foreground emission and point sources in the broad-band imaging.

We have also measured the mass to light ratios of the four central galaxies. 
Each of them retains an associated massive halo.
There is no conclusive evidence to suggest that any of them are more stripped than the others; if anything, the stellar mass of N.1 is marginally lower than expected, compared to both the stellar and dark matter measurements of velocity dispersion.
This is the opposite of behaviour expected if the dark matter associated with N.1 is being stripped.

We have not yet attempted to measure any truncation of the galaxy halos, like \cite{nat09}.
That measurement will be improved by combining our current measurements of strong lensing with spatially extended measurements of weak lensing and flexion currently in preparation, plus multi-object spectroscopic data of member galaxies outside the cluster core. 

\section{Conclusions} \label{sec:conc}

We have presented new {\sl Hubble Space Telescope} imaging and {\sl Very Large Telescope} integrated field spectroscopy of galaxy cluster Abell 3827.
This is a uniquely interesting system for two reasons.
First, it contains an unusually corrugated light distribution, with four almost equally bright central galaxies (not five, as previously thought, because one is a foreground star) within the central 10\,kpc radius.
Second, a gravitationally lensed image of a complex spiral galaxy is fortuitously threaded between the galaxies, allowing the distribution of their total mass to be mapped.
Because of the cluster's $z\!\sim\!0.1$ proximity, this can be achieved with high spatial precision.
We expect this data will be useful beyond this first paper, for several investigations of late-time dark matter dynamics in the poorly-studied (yet theoretically contentious) regime of cluster cores.

We have investigated the possible stripping or deceleration of dark matter associated with the infalling galaxies.
Most interestingly, combining two independent mass mapping algorithms, we find a $1.62^{+0.47}_{-0.49}$\,kpc offset (\ie\ $3.3\sigma$ significance) between total mass and luminous mass in the best-constrained galaxy, including statistical error and sources of systematic error related to the data analysis.
Such an offset does not exist in isolated field galaxies (or it would have been easily detected via strong lensing of quasars).
If interpreted in terms of a drag force caused by weak self-interactions between dark matter particles, this suggests a particle cross-section $\sigma/m\sim(1.7\pm0.7)\times10^{-4}$\,cm$^2$/g.
However, the small absolute offset $<\!2$\,kpc might be caused by astrophysical effects such as dynamical friction, and it is difficult to conclude definitively that real dark matter is behaving differently to CDM.
Detailed hydrodynamical simulations of galaxy infall, incorporating dark matter physics beyond the standard model, are needed to predict its behaviour within a cluster environment, and to more accurately interpret high precision observations.

\section*{Acknowledgments}

The authors are pleased to thank Jay Anderson for advice with CTI correction for {\sl HST/WFC3}, Jean-Paul Kneib for advice using {\sc Lenstool}, and the anonymous referee whose suggestions improved the manuscript.
RM and TDK are supported by Royal Society University Research Fellowships.
This work was supported by the Science and Technology Facilities Council (grant numbers ST/L00075X/1, ST/H005234/1 and ST/I001573/1) and the Leverhulme Trust (grant number PLP-2011-003).
This research was carried out in part at the Jet Propulsion Laboratory, California Institute of Technology, under a contract with NASA.

\noindent {\it Facilities:}
This paper uses data from observations GO-12817 (PI:~R.\,Massey) with the NASA/ESA {\sl Hubble Space Telescope}, obtained at the Space Telescope Science Institute, which is operated by AURA Inc, under NASA contract NAS 5-26555.
This paper also uses data from observations made with ESO Telescopes at the La Silla Paranal Observatory under programmes 093.A-0237 and 294.A-5014 (PI:~R.\,Massey).
We thank the Director General for granting discretionary time, and Paranal Science Operations for running the observations.
The {\sc Lenstool} analysis used the DiRAC Data Centric system at Durham University, operated by the Institute for Computational Cosmology on behalf of the STFC DiRAC HPC Facility (\url{www.dirac.ac.uk}).
This equipment was funded by BIS National E-infrastructure capital grant ST/K00042X/1, STFC capital grant ST/H008519/1, and STFC DiRAC Operations grant ST/K003267/1 and Durham University. DiRAC is part of the National e-Infrastructure. 
LLRW would like to acknowledge the Minnesota Supercomputing Institute, without whose computational support
{\sc Grale} work would not have been possible.

\appendix

\section{Comparison of VIMOS and MUSE spectroscopy} \label{sec:vimosmuse}

\begin{figure}
\begin{center}
\includegraphics[width=0.47\textwidth]{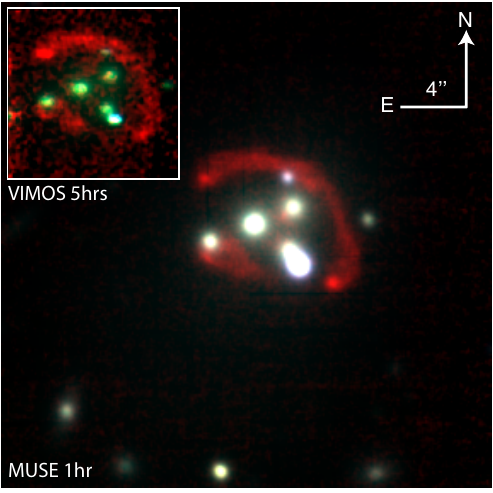}
\caption{
False-colour images of the cluster core, manufactured from the {\sl VLT}/{\sl VIMOS} and {\sl VLT}/{\sl MUSE} IFU 3D data cubes.
The displayed region shows the full field of view of each instrument.
Red, green and blue channels respectively correspond to a narrow-band [{\sc Oii}] image (created as described in section~\ref{sec:slids}), Johnson I band and Johnson V band. }
\label{fig:vimos_vs_muse}
\end{center}
\end{figure}

Our IFU observations of the same (faint and highly elongated) source with both {\sl VLT}/{\sl VIMOS} \citep{vimos} and {\sl VLT}/{\sl MUSE} \citep{muse} provide an early opportunity to compare the practical performance of {\sl VLT}'s old and new spectrographs.
As is apparent in figure~\ref{fig:vimos_vs_muse}, the throughput of {\sl MUSE} is vastly greater, achieving higher signal to noise in a much shorter exposure time.
Crucially for our purposes, the improved flat field stability allows much cleaner foreground subtraction, revealing candidates for the demagnified fifth image near foreground galaxies N.3 and N.4.
{\sl MUSE}'s higher spectral resolution and larger field of view are also responsible for two fortuitous discoveries: our identification of line emission in lensed galaxy A as the [{\sc Oii}] doublet, which we had previously mistaken for H$\alpha$, and the identification of galaxy B as an edge-on spiral, which we had previously mistaken for a strong gravitational lensing arc.
Availability of the new {\sl MUSE} spectrograph has thus revolutionised our interpretation of this cluster.

\section{Alternative lensed image associations} \label{sec:altconfigs}

\begin{figure}
\begin{center}
\includegraphics[width=0.235\textwidth]{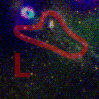}
\includegraphics[width=0.235\textwidth]{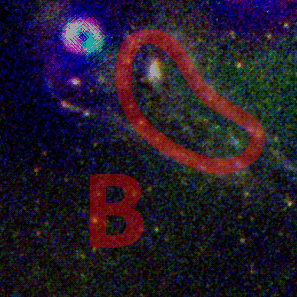}\\
\caption{
Two alternative configurations of lensed image associations near cluster galaxy N.1.
There is a surprising high density of point sources, so it is not obvious which sources correspond to which lensed images.
Configuration L is a small perturbation of the fiducial model in the main paper.
Configuration B is based upon {\sc Lentool}'s best-fit mass model if no identifications are initially made in this region.
However, it is at odds with narrow-band imaging of the source, created from line emission in our {\sl VLT} IFU data (see figure~\ref{fig:imagesub}).
\label{fig:altconfigs}}
\end{center}
\end{figure}

To draw robust conclusions about any offsets between optical emission and associated dark matter, one needs to account for all sources of uncertainty in the mass models.
Statistical uncertainties in the location of the mass peaks around galaxy N.1 are easily obtained. 
There are two types of additional uncertainties: those arising from the mass reconstruction methods, and those from the image identification. 
We explore the latter in this appendix, by considering two alternative image assignments (see figure~\ref{fig:altconfigs}).

In the body of this paper, we presented a fiducial model using what we believe to be the correct associations -- based on the surface brightness, colours and sizes of the knots in {\sc HST} imaging, and the extended distribution of [{\sc Oii}] emission in IFU spectroscopy. 
The only questionable region is around N.1; the other image assignments are secure (though the exact locations of some faint extended images, like Ae and Af, are uncertain to within a few tenths of an arcsecond). 
All the image assignment schemes we have tried produce a common elongation of the mass distribution along the northwest to southeast axis, and a statistically significant offset between the light and mass of N.1, using both {\sc Grale} and {\sc Lenstool}.
However, the amount and the morphology of the offset differs.

\begin{table}
 \centering
  \caption{Changes to the assignments of multiply imaged systems in table~\ref{tab:arcs}, to produce alternative configuration B. Columns denote the ID and position of the image, its major and minor axis, and the angle of the major axis on the sky, anticlockwise from west.}
  \label{tab:altarcs}
  \begin{tabular}{lllllr}
  \hline
  \hline
Name & RA & Dec &  \!\!Major & \!\!Minor & \!\!\!Angle \\
\hline
Aa.4 & $330.47447$ & $-59.946180$ & $0\farcs13$ & $0\farcs09$ & 90$^\circ$ \\
Aa.5 & --- & --- & --- & --- & --- \\
Aa.6 & --- & --- & --- & --- & --- \\
Ab.4 & $330.47452$ & $-59.946347$ & $0\farcs10$ & $0\farcs07$ & 90$^\circ$ \\
Ac.4 & $330.47440$ & $-59.946032$ & $0\farcs07$ & $0\farcs06$ & 70$^\circ$ \\
\hline
\hline
  \end{tabular}
\end{table}

\begin{table*}
 \centering
  \caption{Parameters of the best-fit mass model constructed by {\sc Lenstool}, using alternative associations between lensed images. 
  Positions are relative to the peak of light emission, except for the cluster-scale halo, whose position is relative to the peak of emission from galaxy N.1.
  Errors show 68\% statistical confidence limits due to uncertainty in the lensed image positions.
  \label{tab:altpots}}
  \begin{tabular}{lr@{}lr@{}lr@{}lr@{}lr@{}lr@{}lccc}
  \hline
  \hline
~ & \multicolumn{2}{l}{$\sigma_\mathrm{v}\,$[km/s]\!\!\!} & \multicolumn{2}{c}{$x\,$[\arcsec]} & \multicolumn{2}{c}{$y\,$[\arcsec]} & \multicolumn{2}{c}{$e$} & \multicolumn{2}{c}{$\theta\,$[$^\circ$]} & \multicolumn{2}{c}{$r_\mathrm{core}\,$[\arcsec]} & $\langle\mathrm{rms}_i\rangle\,$[\arcsec] & $\chi^2\!/\mathrm{dof}$ & $\log_{10}{\!(E)}$ \\
\hline
\multicolumn{13}{l}{Alternative configuration L:} & $0.26$ & $50.8/23$ & $-27.8$ \\
N.1 & $185$ & $^{+10}_{-11}$ & $-0.43$ & $^{+0.17}_{-0.16}$ & $-0.69$ & $^{+0.18}_{-0.19}$ & $0.34$ & $^{+0.10}_{-0.14}$ & $50$ & $^{+56}_{-13}$ & \multicolumn{2}{c}{[$\to$0]} \\
N.2 & $187$ & $^{+42}_{-17}$ & $-0.86$ & $^{+0.47}_{-0.33}$ & $-0.40$ & $^{+0.30}_{-0.22}$ & $0.44$ & $^{+0.09}_{-0.13}$ & $176$ & $^{+138}_{-4}$ & \multicolumn{2}{c}{[$\to$0]} \\
N.3 & $241$ & $^{+14}_{-19}$ & $-0.01$ & $^{+0.36}_{-0.25}$ & $-0.23$ & $^{+0.28}_{-0.30}$ & $0.35$ & $^{+0.07}_{-0.11}$ & $28$ & $^{+12}_{-12}$ & \multicolumn{2}{c}{[$\to$0]} \\
N.4 & $261$ & $^{+25}_{-19}$ & $-1.39$ & $^{+0.64}_{-0.21}$ & $0.20$ & $^{+0.38}_{-0.30}$ & $0.13$ & $^{+0.11}_{-0.13}$ & $91$ & $^{+57}_{-14}$ & \multicolumn{2}{c}{[$\to$0]} \\
N.6 & $21$ & $^{+31}_{-11}$ & \multicolumn{2}{c}{[0]} & \multicolumn{2}{c}{[0]} & \multicolumn{2}{c}{[0]} & \multicolumn{2}{c}{[0]} & \multicolumn{2}{c}{[$\to$0]} \\
Cluster & $711$ & $^{+79}_{-85}$ & $2.97$ & $^{+1.91}_{-0.90}$ & $1.50$ & $^{+1.45}_{-1.04}$ & $0.58$ & $^{+0.08}_{-0.14}$ & $70$ & $^{+6}_{-3}$ & $34.52$ & $^{+4.11}_{-4.17}$ \\
\hline
\multicolumn{13}{l}{Alternative configuration B:} & $0.23$ & $34.5/19$ & $-21.3$ \\
N.1 & $252$ & $^{+19}_{-19}$ & $-3.14$ & $^{+0.86}_{-0.78}$ & $-1.17$ & $^{+0.43}_{-0.40}$ & $0.43$ & $^{+0.08}_{-0.17}$ & $150$ & $^{+98}_{-17}$ & \multicolumn{2}{c}{[$\to$0]} \\
N.2 & $216$ & $^{+20}_{-18}$ & $-0.65$ & $^{+0.24}_{-0.40}$ & $0.44$ & $^{+0.40}_{-0.37}$ & $0.40$ & $^{+0.13}_{-0.10}$ & $133$ & $^{+18}_{-12}$ & \multicolumn{2}{c}{[$\to$0]} \\
N.3 & $236$ & $^{+14}_{-26}$ & $-0.15$ & $^{+0.34}_{-0.28}$ & $-0.44$ & $^{+0.32}_{-0.34}$ & $0.43$ & $^{+0.02}_{-0.09}$ & $24$ & $^{+12}_{-9}$ & \multicolumn{2}{c}{[$\to$0]} \\
N.4 & $252$ & $^{+16}_{-21}$ & $-0.63$ & $^{+0.32}_{-0.29}$ & $0.97$ & $^{+0.36}_{-0.42}$ & $0.38$ & $^{+0.08}_{-0.10}$ & $80$ & $^{+15}_{-35}$ & \multicolumn{2}{c}{[$\to$0]} \\
N.6 & $34$ & $^{+54}_{-4}$ & \multicolumn{2}{c}{[0]} & \multicolumn{2}{c}{[0]} & \multicolumn{2}{c}{[0]} & \multicolumn{2}{c}{[0]} & \multicolumn{2}{c}{[$\to$0]} \\
Cluster & $687$ & $^{+63}_{-94}$ & $7.66$ & $^{+1.94}_{-0.81}$ & $2.66$ & $^{+2.12}_{-1.70}$ & $0.70$ & $^{+0.18}_{-0.09}$ & $69$ & $^{+4}_{-7}$ & $38.89$ & $^{+4.41}_{-5.01}$ \\
\hline\hline
  \end{tabular}
\end{table*}

\begin{figure}
\begin{center}
\includegraphics[trim = 12mm 60mm 12mm 34mm, clip, width=0.45\textwidth]{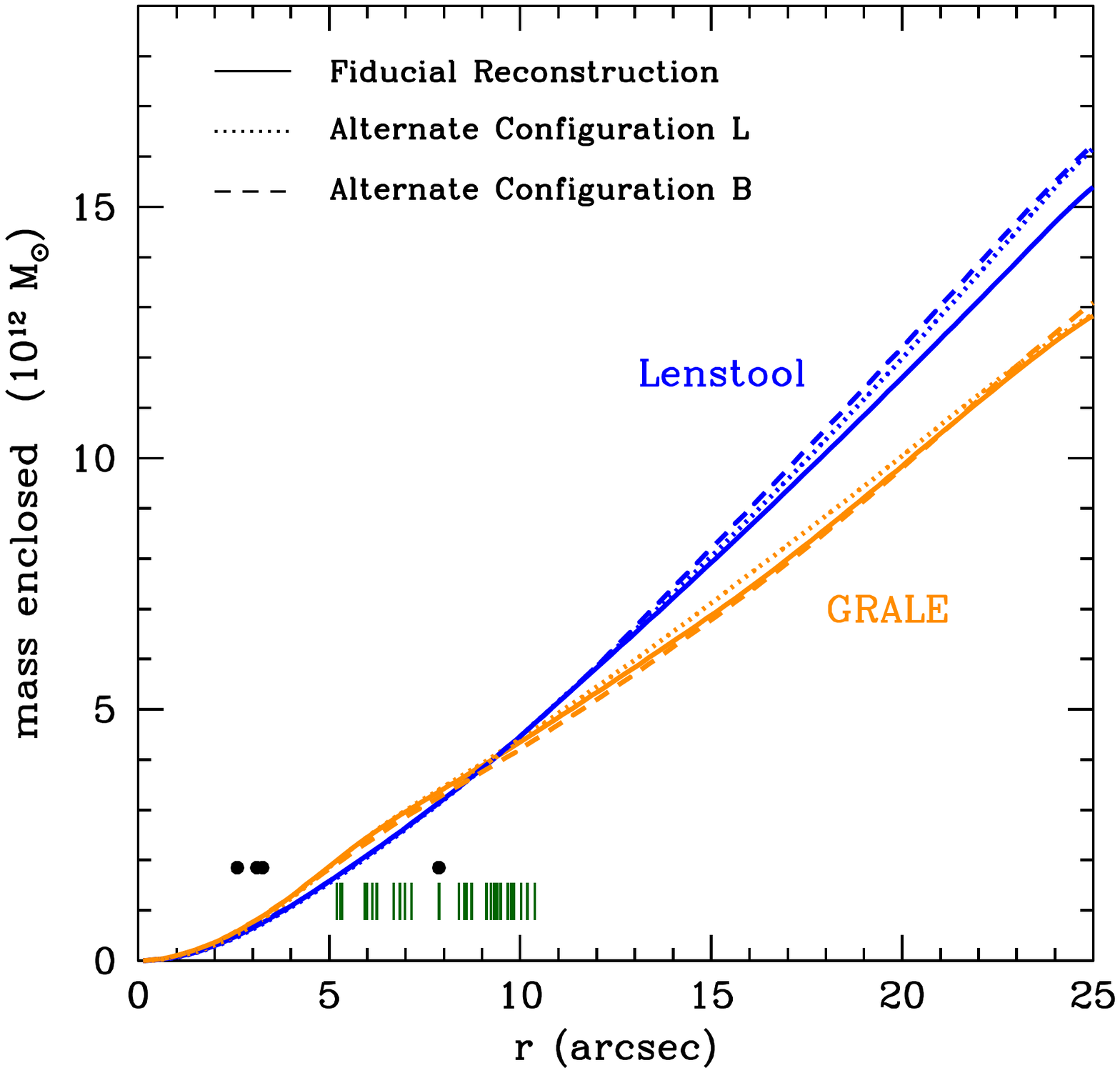}
\caption{
Total mass profile of the cluster core as a function of projected distance from the peak of the {\sc Grale} mass map, for the fiducial model versus the alternate configurations of lensed image associations.
Circles along the bottom show the radial positions of cluster member galaxies N.1--4, and vertical lines mark the positions of lensed images.}
\label{fig:massenc}
\end{center}
\end{figure}

\begin{figure}
\begin{center}
\includegraphics[trim = 8mm 72mm 8mm 37mm, clip, width=0.418\textwidth]{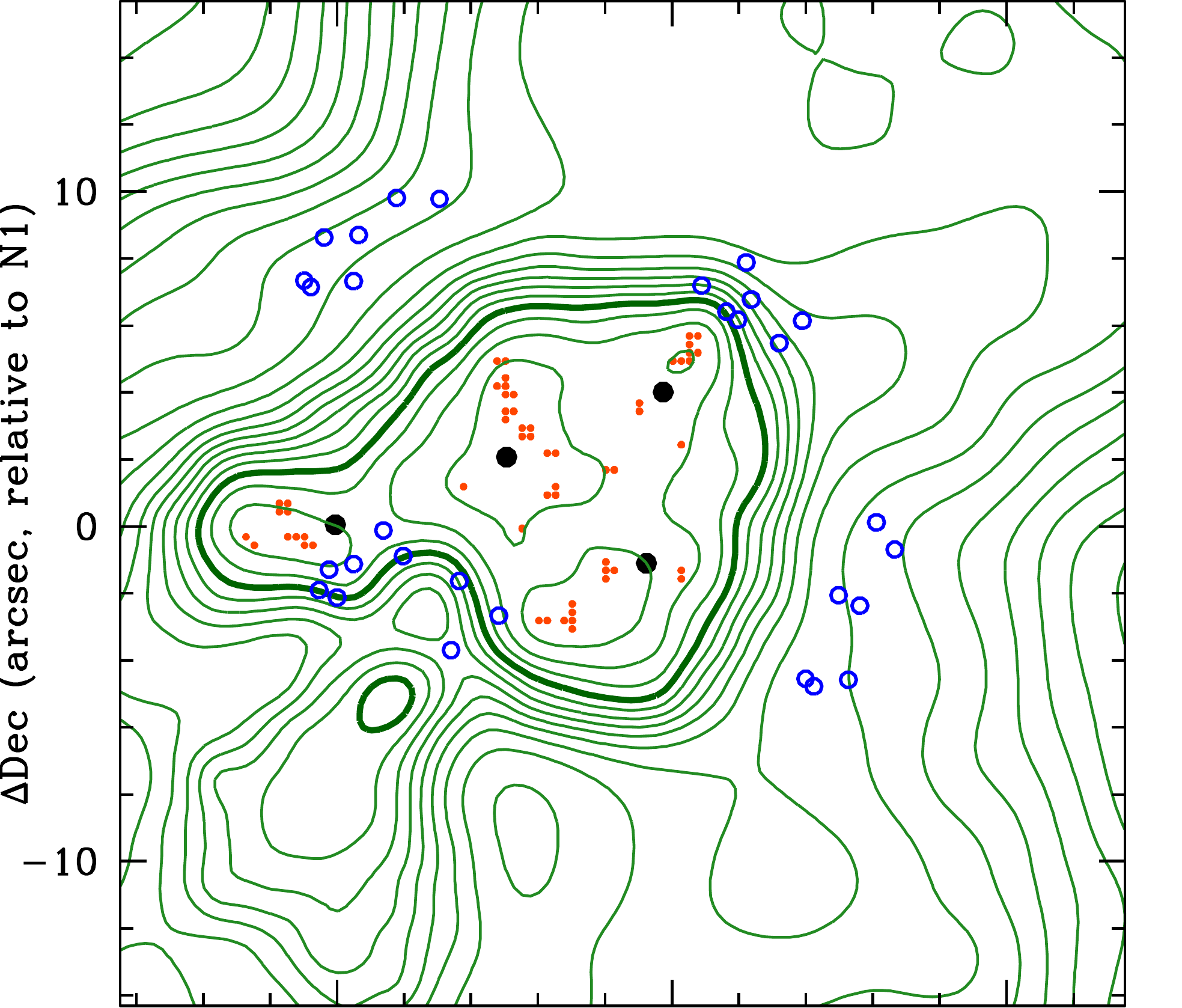}\\
\includegraphics[trim = 8mm 72mm 8mm 37mm, clip, width=0.418\textwidth]{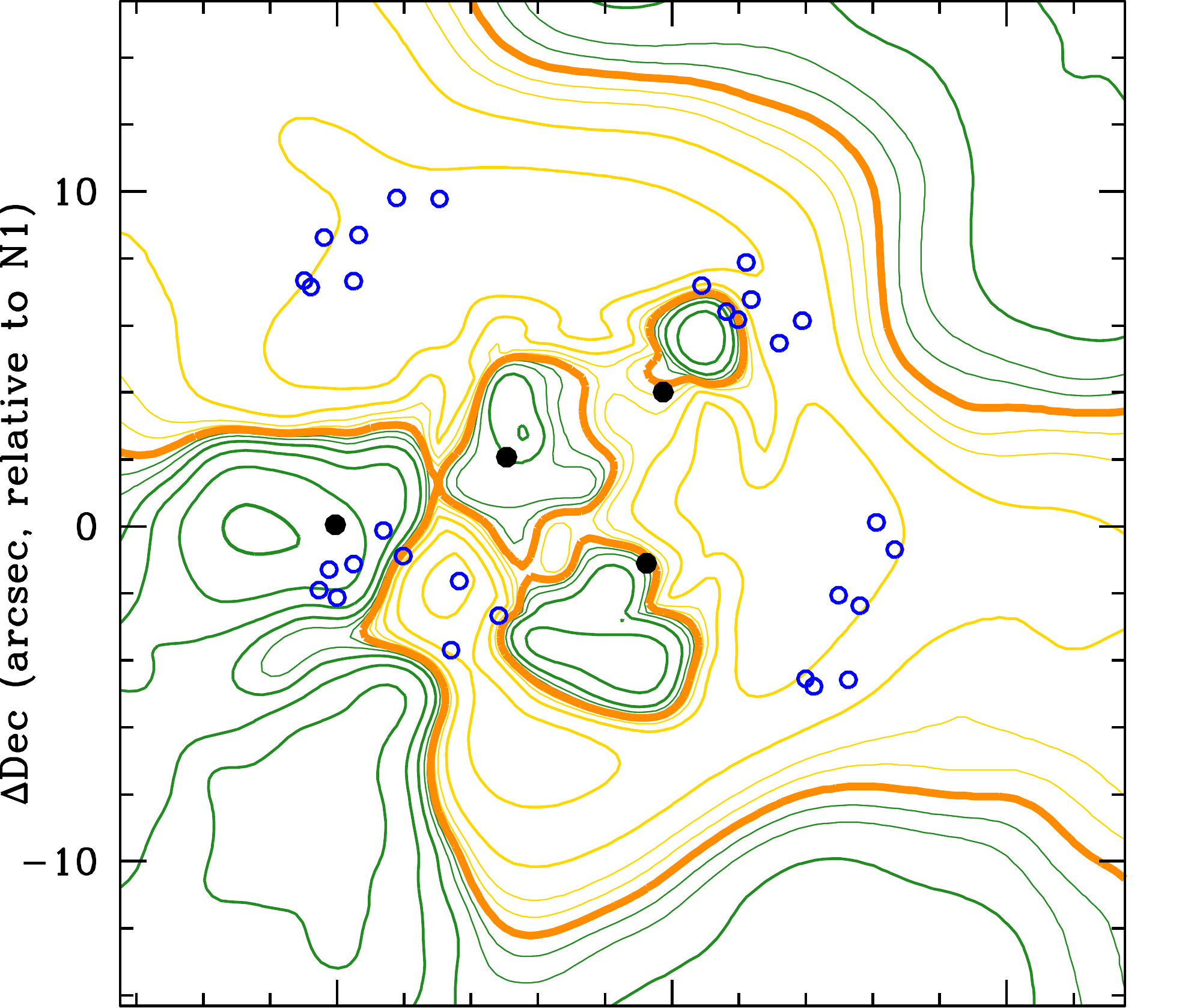}\\
\includegraphics[trim = 8mm 54mm 8mm 37mm, clip, width=0.418\textwidth]{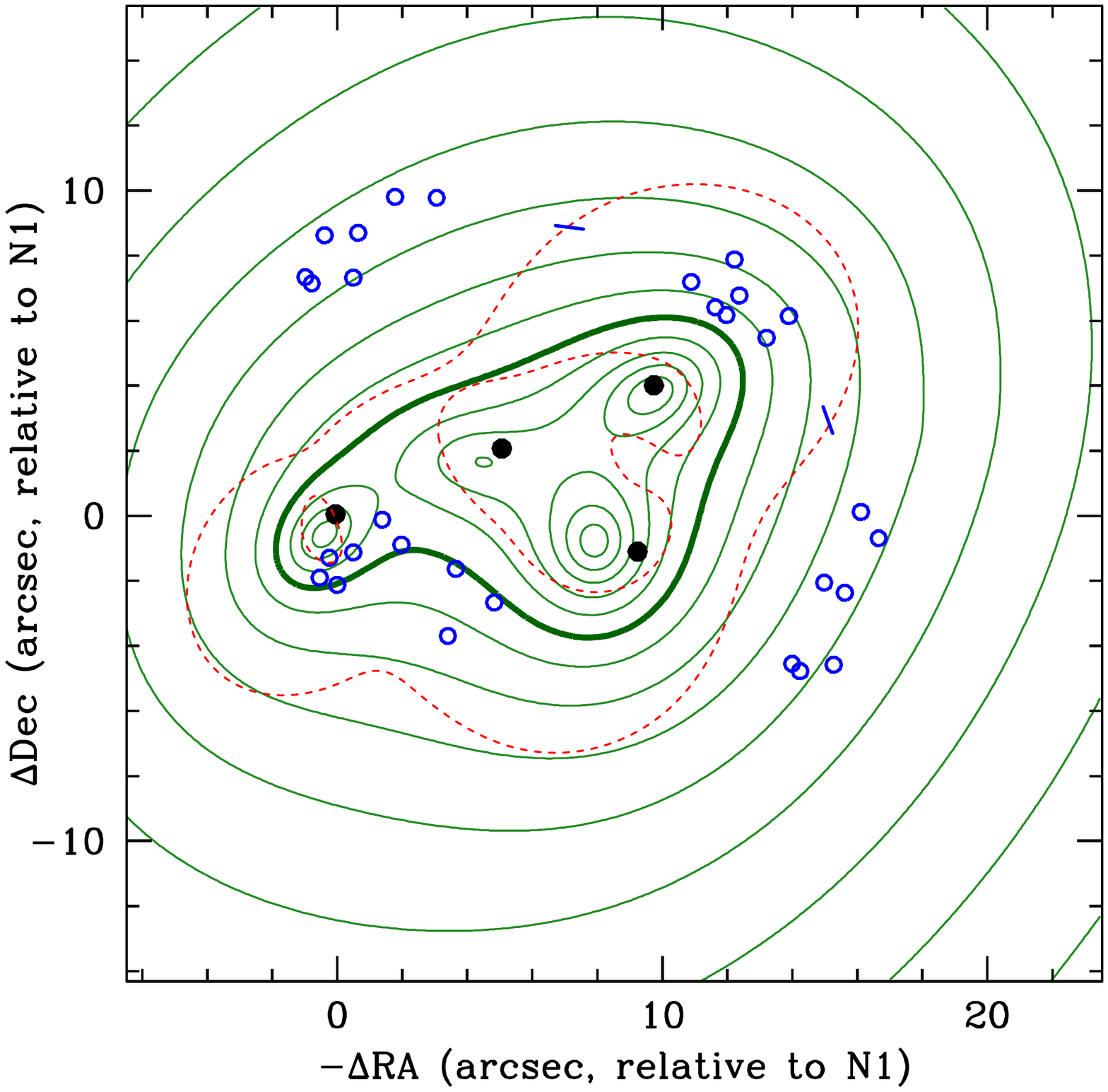}\\
\caption{
Map of total mass in the cluster core, as in figure~\ref{fig:configAGrale} but using Alternate Configuration L to associate lensed images.}
\label{fig:altconfigL}
\end{center}
\end{figure}

Alternative configuration L represents a minor perturbation of the strong lens image assignments, with Aa.6 switched to a fifth image of star formation knot Ab.5.
In the fiducial configuration, {\sc Grale} reproduces the three images of knot Aa, but merged into a continuous arc, which is not consistent with their appearance on HST images.
In configuration L, {\sc Grale} finds a mass $4.44\times10^{12}\,M_\odot$ within $10\arcsec$ of the peak (see figure~\ref{fig:massenc}), which is better able to reproduce the observed distribution of luminous sources near N.1.
However, it introduces an extra mass clump $\sim6\arcsec$ south of N.1 that has no identifiable optical counterpart (see figure~\ref{fig:altconfigL}).
With this configuration, {\sc Lenstool} produces inconsistent predictions of the multiplicity of knot a, and an overall fit marginally worse than the fiducial model.
In the best-fit model, the total mass is $4.48\times10^{12}\,M_\odot$ within the same aperture as above.
The positions of the mass peaks move only within statistical errors compared to the fiducial model (see table~\ref{tab:altpots}), for this and similar minor perturbations to the image assignments.

\begin{figure}
\begin{center}
\includegraphics[trim = 8mm 72mm 8mm 37mm, clip, width=0.418\textwidth]{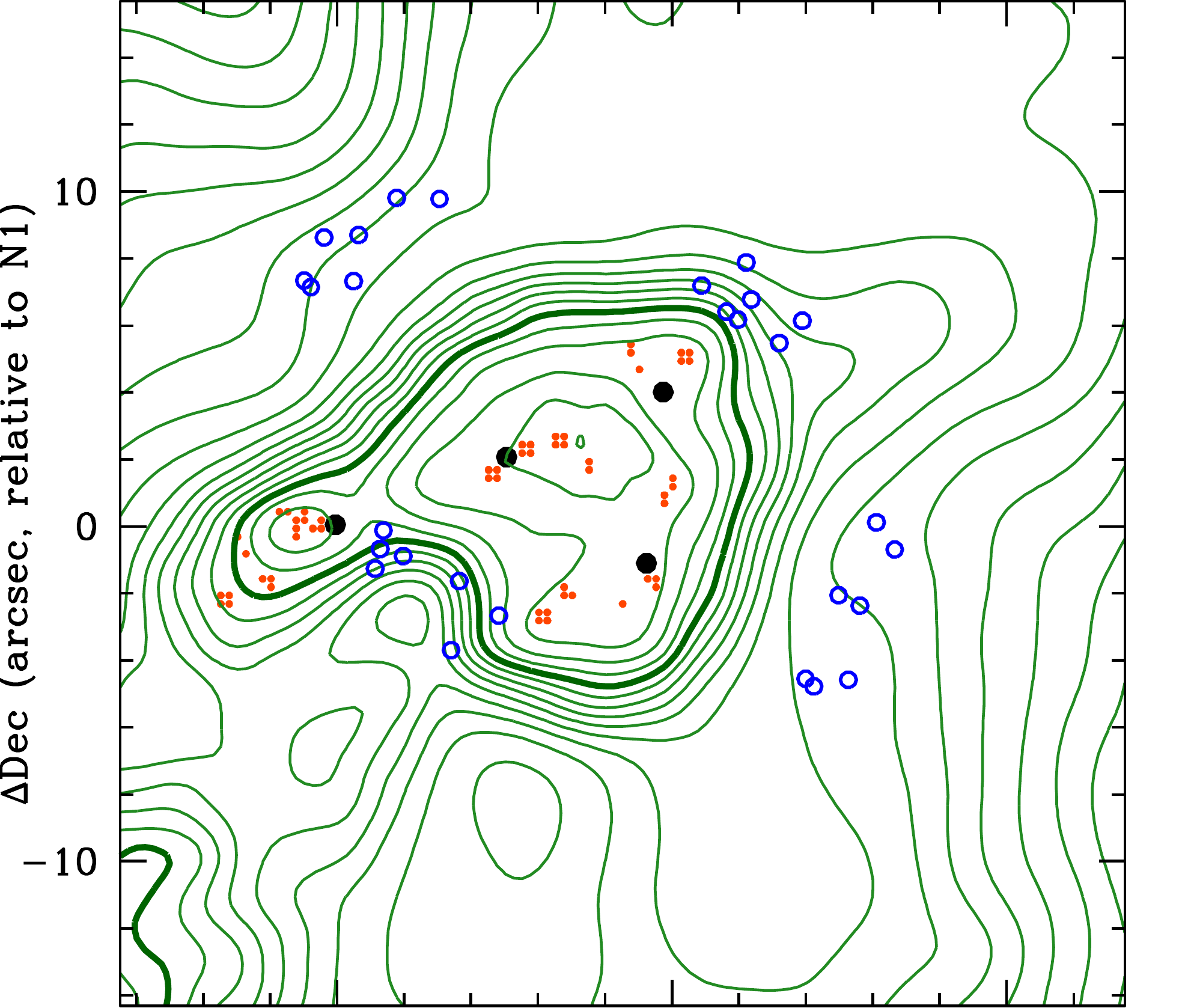}\\
\includegraphics[trim = 8mm 72mm 8mm 37mm, clip, width=0.418\textwidth]{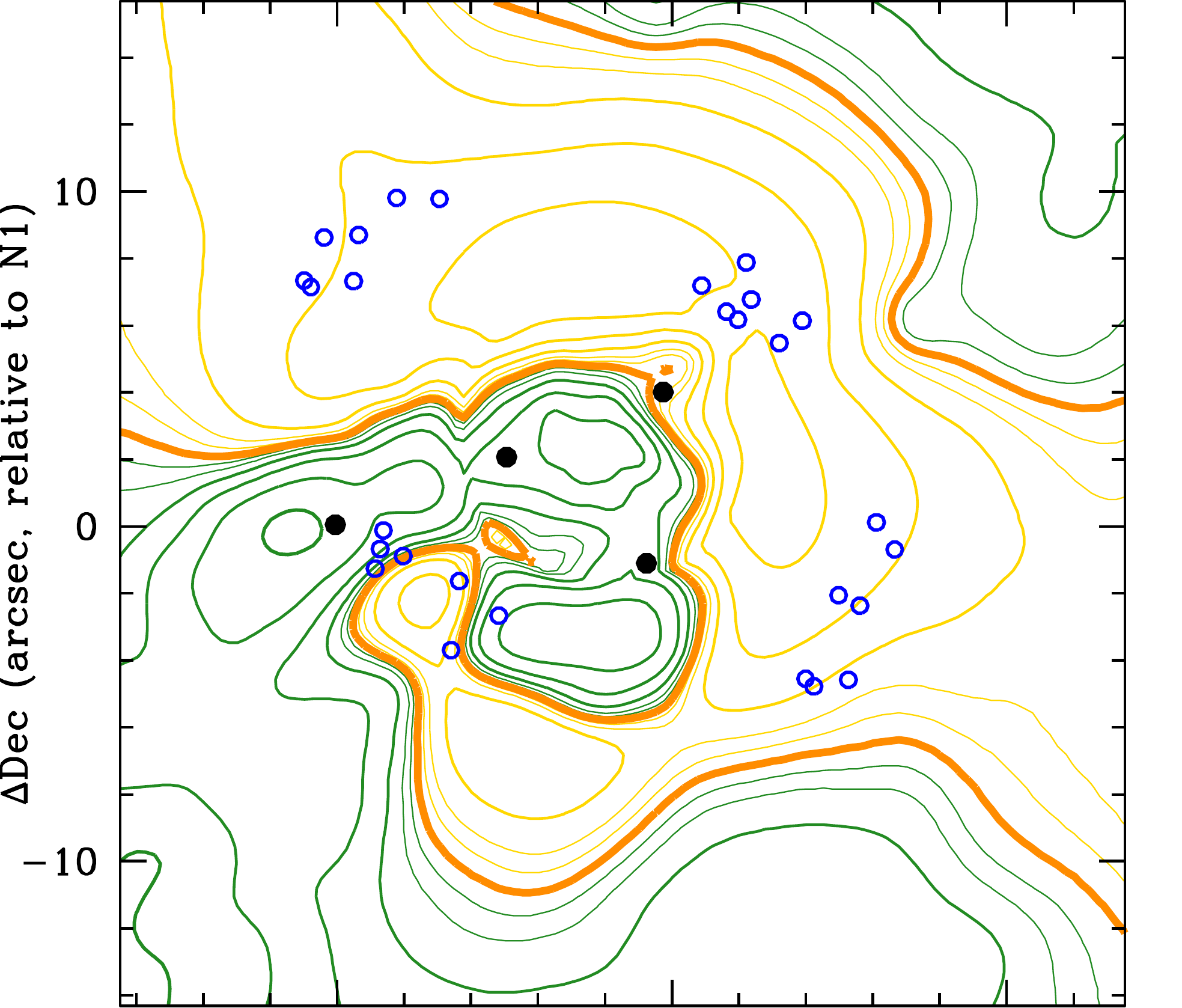}\\
\includegraphics[trim = 8mm 54mm 8mm 37mm, clip, width=0.418\textwidth]{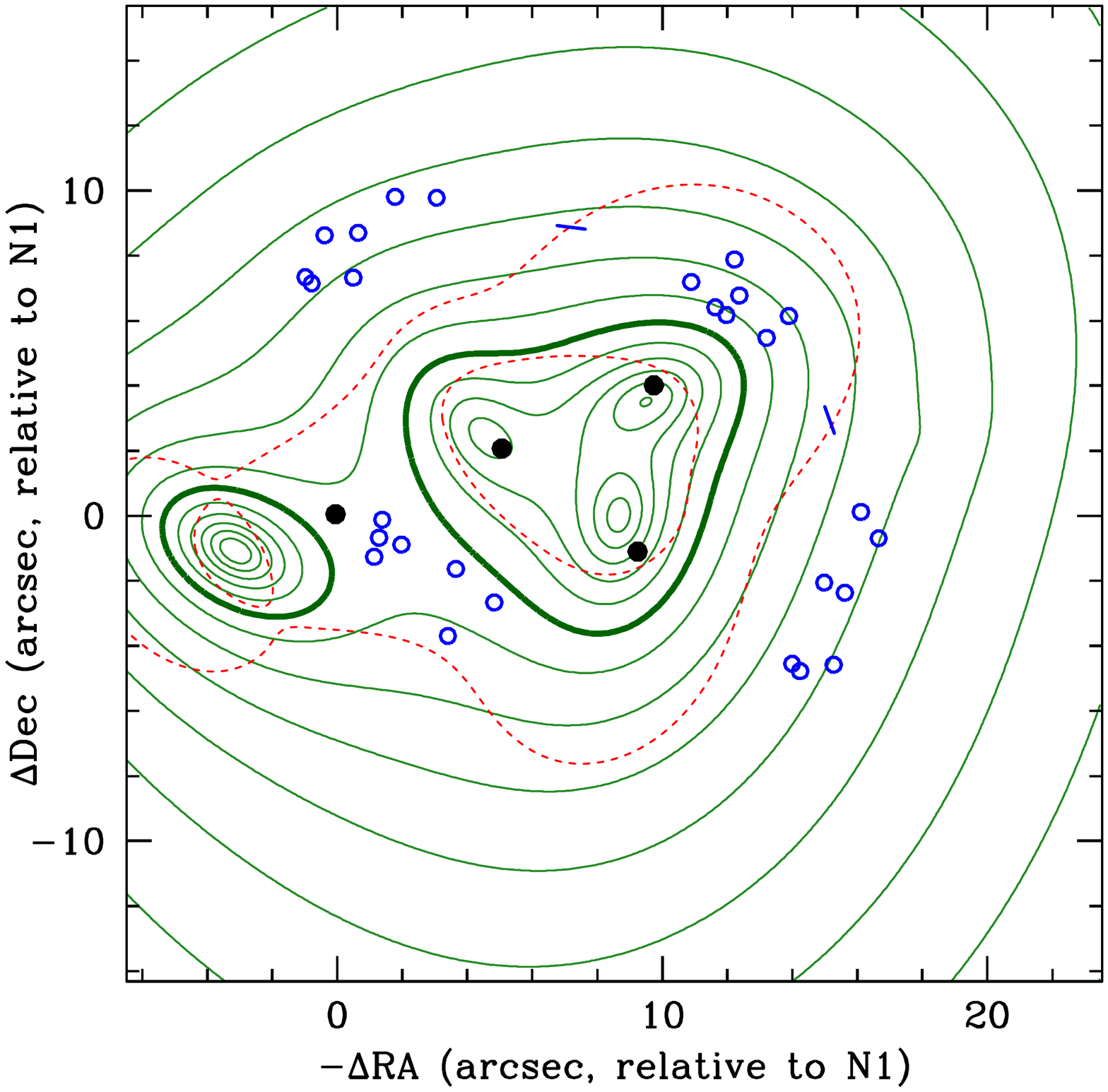}
\caption{
Map of total mass in the cluster core, as in figure~\ref{fig:configAGrale} but using Alternate Configuration B to associate lensed images.
\label{lastpage}}
\label{fig:altconfigB}
\end{center}
\end{figure}

Alternative configuration B had been our preferred configuration before we obtained any {\sl VLT} integrated field spectroscopy.
If star formation knots Aa.4--6, Ab.4 and Ac.4 are left unassigned, and the fit is optimised without any constraints in that region, the best-fit model converges to an assignment like that in table~\ref{tab:altarcs}, reflecting a morphology illustrated in figure~\ref{fig:altconfigs}.
The images immediately south of N.1 in the broad-band images are unaccounted for as strong lens sources, but point sources are frequent in this region, so they could be a chance superposition.
Using these assignments, {\sc Grale} finds a total mass $6.06\times10^{12}\,M_\odot$ within $10\arcsec$ of the peak.
The mass near N.1 becomes a `tail' extending to the southeast (see figure~\ref{fig:altconfigB}).
This is particularly interesting because in some models of particle physics \citep{kah14}, non-zero self-interaction cross-section would indeed disperse the dark matter, rather than simply offsetting it from the light.
A similar extension would also probably be expected in the case of dynamical fraction.
Using configuration B, the best-fit {\sc Lenstool} model with mass $4.48\times10^{12}\,M_\odot$ is apparently better than the fiducial model, with $\langle\mathrm{rms}_i\rangle\!=\!0.23\arcsec$.

Most notably, configuration B's $3\farcs35^{+0.74}_{-0.82}$ offset between the mass and light of N.1 reproduces the result of \cite{ws11}.
This recovery of the ground-based results is reassuring: the star formation knots near N.1 were not resolved in the ground-based imaging.
However, our IFU data shows this image configuration to be incorrect.
[{\sc Oii}] line emission is observed immediately south of N.1 in figure~\ref{fig:imagesub}, inconsistent with the configuration B morphology illustrated in figure~\ref{fig:altconfigs}.
Note that the observed line emission flux has a large spatial gradient, and is detectable even through the obscuration of N.1 because (according to our fiducial model) the gravitational lensing magnification is 53\% greater at the position of knot Aa.4 than at the position of knot Ad.4.
The change in conclusion from \cite{ws11} to this work thus emerges equally from both our new high resolution {\sl HST} imaging {\em plus} our new {\sl VLT} integrated field spectroscopy, and demonstrates the discriminatory power of the combined observations.

\bsp


\begin{thebibliography}{99}
\bibitem[\protect\citeauthoryear{Anderson \& Bedin}{2010}]{and10} Anderson J.\ \& Bedin L.\ 2010 PASP 122, 1035
\bibitem[\protect\citeauthoryear{Anderson}{2014}]{and14} Anderson J.\ 2014 AAS 22412205A
\bibitem[\protect\citeauthoryear{Bacon et al.}{2010}]{muse} Bacon R.\ et al.\ 2010, SPIE 7735, 773508
\bibitem[\protect\citeauthoryear{Bruzual \& Charlot}{2003}]{bc03} Bruzual G.\ \& Charlot S., 2003, MNRAS 344, 1000
\bibitem[\protect\citeauthoryear{Bah\'e et al.}{2012}]{bah12} Bah\'e Y., McCarthy I., Crain R.\ \& Theuns T.\ 2012 MNRAS 424, 1179
\bibitem[\protect\citeauthoryear{Bartelmann \& Schneider}{2001}]{bs01} Bartelmann M.\ \& Schneider P.\ 2001 Phys.\ Rep.\ 340, 291
\bibitem[\protect\citeauthoryear{Bartelmann}{2010}]{barrev} Bartelmann M.\ 2010 Classical \& Quantum Gravity 27, 233001 
\bibitem[\protect\citeauthoryear{Brada\v{c} et al.}{2006}]{bra06} Brada\v{c} M.\ et al.\ 2006 ApJL 652, 937
\bibitem[\protect\citeauthoryear{Brada\v{c} et al.}{2008}]{bra08} Brada\v{c} M.\ et al.\ 2008 ApJ 687, 959
\bibitem[\protect\citeauthoryear{Carrasco et al.}{2010}]{car10} Carrasco E.\ et al.\ 2010 ApJ 715, 160
\bibitem[\protect\citeauthoryear{Chabrier}{2003}]{cha03} Chabrier G.\ 2003 PASP 115, 763
\bibitem[\protect\citeauthoryear{Clowe et al.}{2004}]{clo04} Clowe D., Gonzalez A.\ \& Markevitch M.\ 2004 ApJ 604, 596
\bibitem[\protect\citeauthoryear{Clowe et al.}{2006}]{clo06} Clowe D., Brada\v{c} M., Gonzalez A., Markevitch M., Randall S., Jones C.\ \& Zaritsky D.\ 2006 ApJL 648, 109
\bibitem[\protect\citeauthoryear{Clowe et al.}{2012}]{clo12} Clowe D.\, Markevitch M., Brada\v{c} M., Gonzalez A., Massey R.\ \& Zaritsky D. 2012 ApJ 758, 128
\bibitem[\protect\citeauthoryear{Coe et al.}{2012}]{coe12} Coe D.\ et al.\ 2012 ApJ 757, 22
\bibitem[\protect\citeauthoryear{Dariush et al.}{2010}]{dar10} Dariush A., Raychaudhury S., Ponman T., Khosroshahi H., Benson A., Bower R.\ \& Pearce F.\ 2010 MNRAS 405, 1873
\bibitem[\protect\citeauthoryear{Davis et al.}{1985}]{dav85} Davis M., Efstathiou G., Frenk C.\ \& White S.\ 1985 ApJ 292, 371
\bibitem[\protect\citeauthoryear{Dawson et al.}{2012}]{daw12} Dawson W.\ et al.\ 2012 ApJL 747, 42
\bibitem[\protect\citeauthoryear{Dejonghe}{1987}]{dejong87} Dejonghe H.\ 1987 MNRAS 224, 13
\bibitem[\protect\citeauthoryear{De Plaa et al.}{2007}]{depla07} De Plaa J., Werner N., Bleeker J., Vink J., Kaastra J.\ \& M\'endez M.\ 2007 A\&A 465, 345
\bibitem[\protect\citeauthoryear{Diemand, Kuhlen \& Madau}{2007}]{die07} Diemand J., Kuhlen M.\ \& Madau P.\ 2007 ApJ 667, 859
\bibitem[\protect\citeauthoryear{Eckert et al.}{2014}]{eck14} Eckert D.\ et al.\ 2014, A\&A 570, 119
\bibitem[\protect\citeauthoryear{El\'{\i}asd{\'o}ttir et~al.}{2007}]{eliasdottir07} El\'{\i}asd{\'o}ttir \'A.\ et al.\ 2007, arXiv:0710.5636
\bibitem[\protect\citeauthoryear{Fruchter \& Hook}{2002}]{drizzle} Fruchter A.\ \& Hook R.\ 2002 PASP 114, 144
\bibitem[\protect\citeauthoryear{Gao et al.}{2004}]{gao04} Gao L., De Lucia G., White S.\ \& Jenkins A.\ 2004 MNRAS 352, L1
\bibitem[\protect\citeauthoryear{Gavazzi et al.}{2007}]{slacs4} Gavazzi R.\ et al.\ 2007 ApJ 667, 176
\bibitem[\protect\citeauthoryear{Gillis et al.}{2013}]{gil13} Gillis B.\ et al.\ 2013 MNRAS 431, 1439
\bibitem[\protect\citeauthoryear{Harvey et al.}{2013}]{har13} Harvey D., Massey R., Kitching T., Taylor A., Jullo E., Kneib J.-P., Tittley E.\ \& Marshall P.\ MNRAS 2013 433,1517
\bibitem[\protect\citeauthoryear{Harvey et al.}{2014}]{har14} Harvey D.\ et al.\ 2014 MNRAS 441, 404
\bibitem[\protect\citeauthoryear{Harvey et al.}{2015}]{har15} Harvey D.\, Massey R., Kitching T., Taylor A.\ \& Tittley E.\ 2015 Science in press
\bibitem[\protect\citeauthoryear{Hoekstra \& Jain}{2008}]{hoerev} Hoekstra H.\ \& Jain B.\ 2008 Ann.\ Rev.\ Nucl.\ Part.\ Sci.\ 58, 99
\bibitem[\protect\citeauthoryear{Jauzac et al.}{2012}]{jau12} Jauzac M.\ et al.\ 2012 MNRAS 426, 3369
\bibitem[\protect\citeauthoryear{Jauzac et al.}{2014a}]{jauzac14a} {Jauzac} M.,  {Cl{\'e}ment} B.,  {Limousin} M.,  {Richard} J.,  {Jullo} E., {Ebeling} H.,  {Atek} H.,  {Kneib} J.-P.,  {Knowles} K.,  {Natarajan} P.,  {Eckert} D.,  {Egami} E.,  {Massey} R.\ \& {Rexroth} M.\ 2014 MNRAS 443,1549
\bibitem[\protect\citeauthoryear{Jee et al.}{2014}]{jee14} Jee M., Hoekstra H., Mahdavi A.\ \& Babul A.\ 2014 ApJ 783, 78
\bibitem[\protect\citeauthoryear{Jullo et al.}{2007}]{Lenstool} Jullo E., Kneib J.-P., Limousin M., El'asd\'ottir \'A., Marshall P.\ \& Verdugo T.\ 2007 NJPh 9, 447
\bibitem[\protect\citeauthoryear{Kahlhoefer et al.}{2014}]{kah14} Kahlhoefer F., Schmidt-Hoberg K., Frandsen M.\ \& Sarkar S. 2014 MNRAS 437, 2865
\bibitem[\protect\citeauthoryear{Kennicutt}{1998}]{ken98} {Kennicutt} R.\ 1998 ARA\&A 36, 189
\bibitem[\protect\citeauthoryear{Kneib et al.}{1996}]{kneib96} {Kneib} J.-P.,  {Ellis} R.,  {Smail} I.,  {Couch} W.\ \& {Sharples} R.\ 1996 ApJ 471, 643
\bibitem[\protect\citeauthoryear{Kneib \& Natarajan}{2011}]{knerev} Kneib J.-P.\ \& Natarajan P.\ 2011 A\&A Rev.\ 19, 47
\bibitem[\protect\citeauthoryear{Koopmans et al.}{2006}]{slacs3} Koopmans L., Treu T., Bolton A., Burles S.\ \& Moustakas L.\ 2006 ApJ 649, 599
\bibitem[\protect\citeauthoryear{Krist, Hook \& Stoehr}{2011}]{tinytim} Krist J., Hook R.\ \& Stoehr F.\ 2011 Proc.\ SPIE vol.\ 8127, 81270J-1
\bibitem[\protect\citeauthoryear{Leauthaud et al.}{2007}]{lea07} Leauthaud A.\ et al.\ 2007 ApJS 172, 219
\bibitem[\protect\citeauthoryear{Le F{\`e}vre et al.}{2003}]{vimos} Le F\`evre O.\ et al.\ 2003, SPIE 4841, 1670
\bibitem[\protect\citeauthoryear{Le F{\`e}vre et al.}{2013}]{lef13} Le F{\`e}vre O.\ et al.\ 2013 A\&A 559, 14
\bibitem[\protect\citeauthoryear{Liesenborgs, De Rijcke \& Dejonghe}{2006}]{lie06} Liesenborgs J., De Rijcke S.\ \& Dejonghe H.\ 2006 MNRAS 367, 1209
\bibitem[\protect\citeauthoryear{Liesenborgs et al.}{2007}]{lie07} Liesenborgs J., De Rijcke S., Dejonghe H.\ \& Bekaert P.\ 2007 MNRAS 380, 1729
\bibitem[\protect\citeauthoryear{Liesenborgs et al.}{2008a}]{lie08a} Liesenborgs J., De Rijcke S., Dejonghe H.\ \& Bekaert P.\ 2008a MNRAS 386, 307
\bibitem[\protect\citeauthoryear{Liesenborgs et al.}{2008b}]{lie08b} Liesenborgs J., De Rijcke S., Dejonghe H.\ \& Bekaert P.\ 2008b MNRAS 389, 415
\bibitem[\protect\citeauthoryear{Liesenborgs et al.}{2009}]{lie09} Liesenborgs J., De Rijcke S., Dejonghe H.\ \& Bekaert P.\ 2009 MNRAS 397, 341
\bibitem[\protect\citeauthoryear{Liesenborgs \& De Rijcke}{2012}]{lie12} Liesenborgs J.\ \& De Rijcke S.\ 2012 MNRAS 425, 1772
\bibitem[\protect\citeauthoryear{{Limousin}, {Kneib} \& {Natarajan}}{{Limousin}  et~al.}{2005}]{limousin05} Limousin M., Kneib J.-P.\ \& Natarajan P., 2005 MNRAS 356, 309
\bibitem[\protect\citeauthoryear{Limousin et al.}{2007}]{lim07} Limousin M., Kneib J.-P., Bardeau S., Natarajan P., Czoske O., Smail I., Ebeling H.\ \& Smith G.\ 2007 A\&A 461, 881
\bibitem[\protect\citeauthoryear{Limousin et al.}{2012}]{lim12} Limousin M.\ et al.\ 2012 A\&A 544, 71
\bibitem[\protect\citeauthoryear{Mahdavi et al.}{2007}]{mah07} Mahdavi A., Hoekstra H., Babul A., Balam D.\ \& Capak P.\ 2007 ApJ 668, 806
\bibitem[\protect\citeauthoryear{Mandelbaum et al.}{2006}]{man06} Mandelbaum R.\, Seljak U., Kauffmann G., Hirata C.\ \& Brinkmann  J.\ 2006 MNRAS 368, 715
\bibitem[\protect\citeauthoryear{Massey et al.}{2010}]{m10} Massey R., Stoughton C., Leauthaud A., Rhodes J., Koekemoer A., Ellis R. \& Shaghoulian E.\ 2010 MNRAS 401, 371
\bibitem[\protect\citeauthoryear{Massey \& Goldberg}{2008}]{wlsl} Massey R.\ \& Goldberg D.\ 2008 ApJ 673, 111
\bibitem[\protect\citeauthoryear{Massey, Kitching \& Richard}{2010}]{mrev} Massey R., Kitching T.\ \& Richard J.\ 2010 Rep.\ Prog.\ Phys.\ 73, 086901 
\bibitem[\protect\citeauthoryear{Massey, Kitching \& Nagai}{2011}]{mkn11} Massey R., Kitching T.\ \& Nagai D.\ 2011 MNRAS 413, 1709
\bibitem[\protect\citeauthoryear{Massey et al.}{2014}]{m14} Massey R.\ et al.\ 2014 MNRAS 439, 887
\bibitem[\protect\citeauthoryear{Merten et al.}{2011}]{mer11} Merten J.\ et al.\ 2011 MNRAS 417, 333
\bibitem[\protect\citeauthoryear{Mohammed et al.}{2014}]{moh14} Mohammed I., Liesenborgs J., Saha P.\ \& Williams L.\ 2014 MNRAS 439, 2651
\bibitem[\protect\citeauthoryear{Nagai et al.}{2005}]{nag05} Nagai D.\ \& Kravtsov A.\ 2005 ApJ 618, 557
\bibitem[\protect\citeauthoryear{Natarajan et al.}{2009}]{nat09} Natarajan P.\ , Kneib J.-P., Smail I., Treu T., Ellis R., Moran S., Limousin M.\ \& Czoske O.\ 2009 ApJ 693, 970
\bibitem[\protect\citeauthoryear{Parker et al.}{2007}]{par07} Parker L., Hoekstra H., Hudson M., van Waerbeke L.\ \& Mellier Y.\ 2007 ApJ 669, 21
\bibitem[\protect\citeauthoryear{Pe\~narrubia, McConnachie \& Navarro}{2008}]{pen08} Pe\~narrubia J., McConnachie A.\ \& Navarro J.\ 2008 ApJ 672, 904
\bibitem[\protect\citeauthoryear{Peng et al.}{2010}]{galfit} Peng C., Ho L., Impey C.\ \& Rix H.-W.\ 2010 AJ 139, 2097
\bibitem[\protect\citeauthoryear{Peter et al.}{2012}]{peterreview} Peter A.\ et al.\ 2012 arXiv:1208.3026
\bibitem[\protect\citeauthoryear{Plummer}{1911}]{plum11} Plummer H.\ 1911 MNRAS 71, 460
\bibitem[\protect\citeauthoryear{Postman et al.}{2012}]{pos12} Postman M.\ et al.\ 2012 ApJ 756, 159
\bibitem[\protect\citeauthoryear{Randall et al.}{2008}]{ran08} Randall S., Markevitch M., Clowe D., Gonzalez A.\ \& Brada\v{c} M.\ 2008 ApJ 679, 1173
\bibitem[\protect\citeauthoryear{Refregier}{2003}]{refrev} Refregier A.\ 2003 ARA\&A 41, 645
\bibitem[\protect\citeauthoryear{Rhodes et al.}{2007}]{rho07} Rhodes J.\ et al.\ 2007 ApJS 172, 203
\bibitem[\protect\citeauthoryear{Roediger et al.}{2014}]{roe14a} Roediger E.\ et al.\ 2014 arXiv:1409.6300
\bibitem[\protect\citeauthoryear{Sabbi et al.}{2009}]{calwf3} Sabbi E.\ et al.\ 2009, WFC3 Instrument Science Report 2009-13
\bibitem[\protect\citeauthoryear{Schaye et al.}{2015}]{eagle} Schaye J.\ et al.\ 2015, MNRAS 446, 521
\bibitem[\protect\citeauthoryear{Smith et al.}{2005}]{smith05} {Smith} G.,  {Kneib} J.-P.,  {Smail} I.,  {Mazzotta} P.,  {Ebeling} H.\ \& {Czoske} O.\ 2005 MNRAS 359, 417
\bibitem[\protect\citeauthoryear{Smith et al.}{2010}]{smi10} Smith G.\ et al.\ 2010 MNRAS 409, 169
\bibitem[\protect\citeauthoryear{Smith et al.}{2012}]{calacs} Smith L.\ et al.\ 2012 AAS 21924101S
\bibitem[\protect\citeauthoryear{Vazedkis}{1999}]{vaz99} Vazdekis A.\ 1999 ApJ 513, 224
\bibitem[\protect\citeauthoryear{Wetzel, Cohn \& White}{2009}]{wet09} Wetzel A., Cohn J.\ \& White M.\ 2009 MNRAS 395, 1376
\bibitem[\protect\citeauthoryear{Williams \& Saha}{2011}]{ws11} Williams L.\ \& Saha P.\ 2011 MNRAS 415, 448
\bibitem[\protect\citeauthoryear{Wu et al.}{2012}]{wu12} Wu H.-Y., Hahn O., Wechsler R., Behroozi P.\ \& Mao Y.-Y.\ 2013 ApJ 767, 23
\bibitem[\protect\citeauthoryear{Young et al.}{2011}]{you11} Young O., Thomas P., Short C.\ \& Pearce F.\ 2011 MNRAS 413, 691

\end{thebibliography}
\end{document}